\documentclass{iopart}

\usepackage{graphicx}
\usepackage[utf8]{inputenc}
\usepackage{wrapfig}
\usepackage{blindtext} 
\usepackage{todonotes}
\usepackage{iopams}  
\usepackage{mathtools}

\usepackage{float}
\usepackage{csquotes}
\usepackage{subfigure}
\usepackage{amsmath}
\usepackage{tipa}
\usepackage[textwidth=6in]{geometry}
\usepackage{multicol}
\usepackage{multirow}
\usepackage[inline]{enumitem}
\usepackage{lipsum}
\usepackage{adjustbox}
\usepackage{natbib}
\usepackage[english]{babel}
\usepackage[breaklinks]{hyperref}
\usepackage[section]{placeins}
\begin{document}
%\setcitestyle{square}

\title[]{Direct mapping from PET coincidence data to proton-dose and positron activity using a deep learning approach}
\author{Atiq~Ur~Rahman$^1$ $^2$,~Mythra~Varun~Nemallapudi$^1$\thanks{Corresponding~author.}, Cheng-Ying~Chou$^3$\thanks{Corresponding~author.}
,~Shih-Chang~Lee$^1$,~Chih-Hsun~Lin$^1$}
%\address{Institute}
\address{$^1$ Institute of Physics, Academia Sinica, Taipei 11529, Taiwan }
\address{$^2$ Department of Physics, National Central University, Taoyuan 320317, Taiwan}
\address{$^3$ Department of Biomechatronics Engineering, National Taiwan University, Taipei 10617, Taiwan}
\ead{\href{mailto:mythravarun01@gmail.com}{mythravarun01@gmail.com}}
\ead{\href{mailto:chengying@ntu.edu.tw}{chengying@ntu.edu.tw}}
\vspace{10pt}
\begin{indented}
\item[] 12 August 2022
\end{indented}
\begin{abstract}
 \textit{Objective}. Obtaining the intrinsic dose distributions in particle therapy is a challenging problem that needs to be addressed by imaging algorithms to take advantage of secondary particle detectors. In this work, we investigate the utility of deep learning methods for achieving direct mapping from detector data to the intrinsic dose distribution.
 \textit{Approach}. We performed Monte Carlo simulations using GATE/Geant4 10.4 simulation toolkits to generate a dataset using human CT phantom irradiated with high-energy protons and imaged with compact in-beam PET for realistic beam delivery in a single-fraction ($\sim$2Gy). We developed a neural network model based on conditional generative adversarial networks to generate dose maps conditioned on coincidence distributions in the detector. The model performance is evaluated by the mean relative error, absolute dose fraction difference, and shift in Bragg peak position.
 \textit{Main results}.  The relative deviation in the dose and range of the distributions predicted by the model from the true values for  mono-energetic irradiation between 50 MeV and 122 MeV lie within 1\% and 2\%, respectively. This was achieved using $\mathrm{10^5}$ coincidences acquired five minutes after irradiation.
The relative deviation in the dose and range for spread-out Bragg peak distributions were within 1\% and 2.6\% uncertainties, respectively.   
\textit{Significance}. An important aspect of this study is the demonstration of a method for direct mapping from detector counts to dose domain using the low count data of compact detectors suited for practical implementation in particle therapy. Including additional prior information in the future can further expand the scope of our model and also extend its application to other areas of medical imaging. 
\end{abstract}
\vspace{2pc}
\section{Introduction}\label{section:intro}
Fast, precise and physically optimized in vivo monitoring for dose and range verification during proton therapy is a challenging and critical task that can expand the scope of treatment planning.   There are two main methods to perform dose and range verification.   The first method requires the measurement of primary or secondary protons using some invasive implantable device or the direct measurement of high-energy proton beams exiting the tumor \citep{lu2008,watts2009proton}.   The scope of the invasive method is limited because it requires implanting a dosimeter in the radiation path of the tumor area, which may not always be possible and applicable; it provides information only at the implanted site.

Alternatively, secondary particles such as positron-annihilation gamma (PAG) and prompt gamma (PG) emitting isotopes generated during proton beam irradiation can also provide dose and range verification information.
The inelastic nuclear interaction of high-energy protons with elemental nuclei results in the production of isotopes with an altered nuclear structure from the parent nucleus represented, e.g. by $^{16}$O(p,pn)$^{15}$O. Positron emitting isotopes are a category of daughter nuclei, referred to as secondary particles, which further undergo a $\beta^+$ emission. The production intensity of such secondaries depends on their energy-dependent production cross sections. The trail of isotopes left by a proton beam follows the dose distribution closely and ceases to exist when the proton energy drops below the threshold value for that isotope's production. This results in a systematic gap from the Bragg peak \citep{refe2,min2006prompt}, \citep{refe1,refe3}.

To obtain an accurate PET distribution \citep{refe4} requires consideration of photon attenuation, detector efficiency, the uncertainty introduced by non-linearities in the positron range, and randomness and scattering within the detector \citep{refe7}.
  For PET image reconstruction, analytical and iterative reconstruction are conventional reconstruction techniques.   Analytic reconstruction techniques are based on non-random conjectures and attempt to estimate images using deterministic mathematical solutions \citep{alessio2005analytical}.   Analytical reconstruction ignores noise contributions, whereas iterative methods can incorporate noise effects and make the final estimate better than the previous ones, with each iteration updating the image \citep{refe8}.
   However, iterative reconstruction methods require an accurate and fast estimation of the system response matrix (SRM), which should include factors such as geometry, noise effects, and local variations like changes in the channel detection efficiency of readouts.   The simulation and integration of all components of SRM make it a large-scale, complex, and computationally intensive task \citep{refe9}. Subsequent to PET activity reconstruction, mapping the reconstructed activity profiles to dose distributions is also complicated in actual patient treatment because of the different physical processes involved in the generation of PAG isotopes compared to the dose deposition \citep{refe10_1}.
 
Two main approaches have been proposed as solutions for validating doses.   The first is an indirect method of verifying the dose-activity profile of a positron emitter.   One of the methods \citep{refe10,refe12} is an interactive solution based on a direct comparison between the simulated and measured signals of ${\beta^+}$ superimposed on a CT image.
If the positions of the two contours do not match, the probable cause of the deviation is speculated, and the PET signal is recalculated after modifications are made to the treatment planning system.   Although the indirect method is satisfactory in accuracy and easy to implement, it cannot directly estimate the difference between delivered and planned doses.   The second method is a direct method in which the dose is plotted from the activity profile of the positron emitter.
An attempt of this type \citep{refe11} was presented, implementing an analytical model to estimate the dose from 3D information preserved in PET signals.   The estimated kernel, the Positron Emitter Species Matrix (PESM), was de-convoluted with the reconstructed positron activity to yield a dose prediction within a 2\% difference from the delivered dose.   Other proposed methods were analytical methods based on modeling functions applied to reconstructed PET activity profiles and have better accuracy than interactive methods \citep{refe14,refe15,refe16} when evaluating homogeneous phantoms.

A pioneering statistical method \citep{refe18} applied Maximum Likelihood Expectation Maximization (MLEM) combined with filtering to obtain a 2D dose distribution, showing promise with mono-energetic and spread-out Bragg peaks (SOBP) under statistical noise.   The relative error of the result was within 10\%. 
\citep{refe19} trained a recurrent neural network (RNN) using simulated depth distributions of $^{11}$C and $^{15}$O to obtain dose maps.
However, the study \citep{refe19} is based on intrinsic isotopic distributions, which are impossible to measure in a patient and therefore cannot be verified experimentally.
The second sequence of the study based on the RNN model combined the PET detector model, the stopping power of the proton beam, and anatomical information using the reconstructed dose mapping profile \citep{refe20}.   Another study \citep{refe20_1} used deep learning methods to predict proton dose in 3D.   The study was carried out on a ring-shaped PET prototype with 88,000 crystal elements to reconstruct PET activity using single-slice rebinning reconstruction and 2D OSEM iterative reconstruction algorithms.   The PET signal-based dose mapping solutions described above are directly or indirectly based on the activity profiles reconstructed using conventional reconstruction methods. 
It is desirable to map the dose distributions directly from the detector data without performing traditional reconstruction, and a neural network can be trained on the detector-induced uncertainties. 

Previously, a deep learning-based solution named DeepPET using an encoder-decoder model was proposed for diagnostic PET systems, which directly transferred the sinogram data into the activity space after training with high-quality real data.   It delivered better image quality than the traditional optimized image reconstruction method  \citep{refe24}.   Another deep learning-based direct reconstruction study was conducted \citep{refe22}, which showed robustness and accuracy compared to conventional reconstruction methods.   Deep learning can be introduced in the dose reconstruction problem after conventional reconstruction of PET activity distribution.   This approach is a deep learning-based post-reconstruction filtering to obtain dose maps that still rely on regular reconstruction, as implemented in Hu \textit{et al}. However, direct proton dose mapping from raw detector data has not yet been explored.
In our work, we propose a deep learning framework to directly map the detector data into the corresponding dose distribution without reconstructing the PET activity distribution.
The same approach can also be used to directly reconstruct the activity profiles of the isotopes by using the 2D raw coincidence projection data as input information.   Previous solutions for in vivo dose monitoring utilized large PET scanners and heavy PG detectors.   Compact PET detectors offer the advantage of being integrated with a treatment gantry and being positioned closer to the patient, which is important for maximizing the available signal and recovering fast-decaying species of positron-emitting isotopes.   Our study aims to develop the neural network for such compact PET detectors described in \citep{refe5,refe5_2}, which are ideal for practical application in the treatment environment where detectors have limited space and maneuverability.  
 Our study's motivation for using a small detector was to examine the possibility of using a miniature imaging solution combined with the feature extraction capabilities of deep networks as an in vivo monitoring solution.   A quantitative study was also conducted to estimate the average number of detector counts required to achieve acceptable accuracy using the palm-sized detector prototype.

The state-of-the-art deep generative models are variant autoencoder (VAE) and GAN models, used extensively  for image synthesis problems \citep{oussidi2018deep}.   GAN models are superior to VAEs because of their ability to extract features from complex data and generate high-quality images.   GANs are a good choice because of their excellent ability to learn domain-dependent and domain-independent features.   In our study, we used a deep generative model initially proposed \citep{MIRZA} as a conditional GAN network and first implemented in an image-to-image translation problem \citep{refe23}. Zhang \textit{et al}., also applied the disocGAN variant of generative adversarial networks (GAN), which is close to the model used in our research.   However, there are significant differences between our study and their disocGAN-based study.   We used only 1,024 detector crystal elements and raw coincidence detector data without using image reconstruction algorithms or filters.   Our work is supervised learning, which requires labeled pairs of source and target images as training data  \citep{refe23,Lin_2018_CVPR}. 
In generative models, once the generator and the discriminator are alternately trained, the trained generator model is obtained and used as an independent model for prediction.   In our work, dose maps can be obtained as output by importing 2D coincidence projection data into the model.   By applying this model, we can train the neural network for the underlying imaging physics, detector geometry, and detector non-uniformity directly from experimental data, thus making the predictions more reliable.
\section{Materials and methods}
\label{section:materialsmethod}
\subsection{Simulation setup for proton beam irradiation}\label{subsection:beamSimulation}
\begin{figure}[ht]
\center
\subfigure[]{\label{fig:Dicom_dose}\includegraphics[trim={0cm .5cm 0cm 0cm},clip,scale=.57]{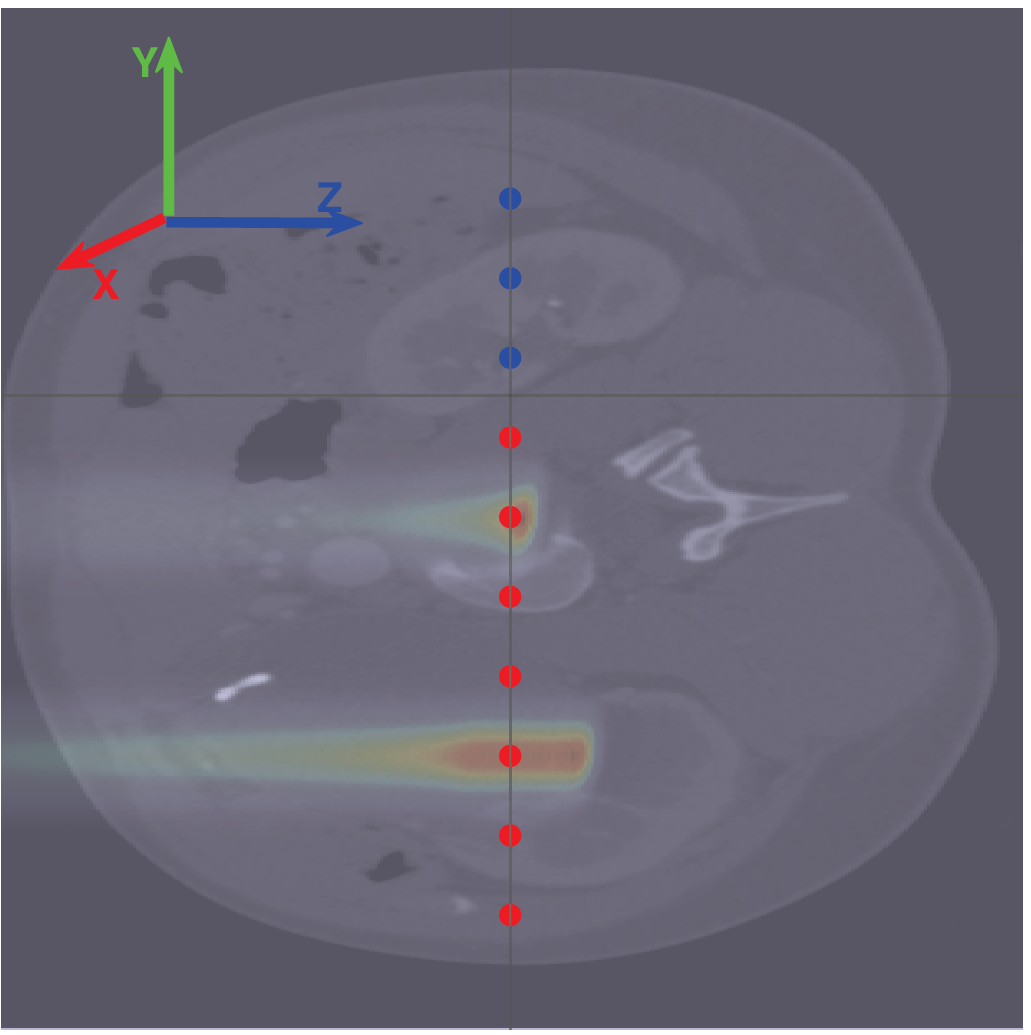}}
\vspace*{1cm}
\subfigure[]{\label{fig:DET_phantom}\includegraphics[trim={0cm 0cm 0cm 0cm},clip,scale=.67]{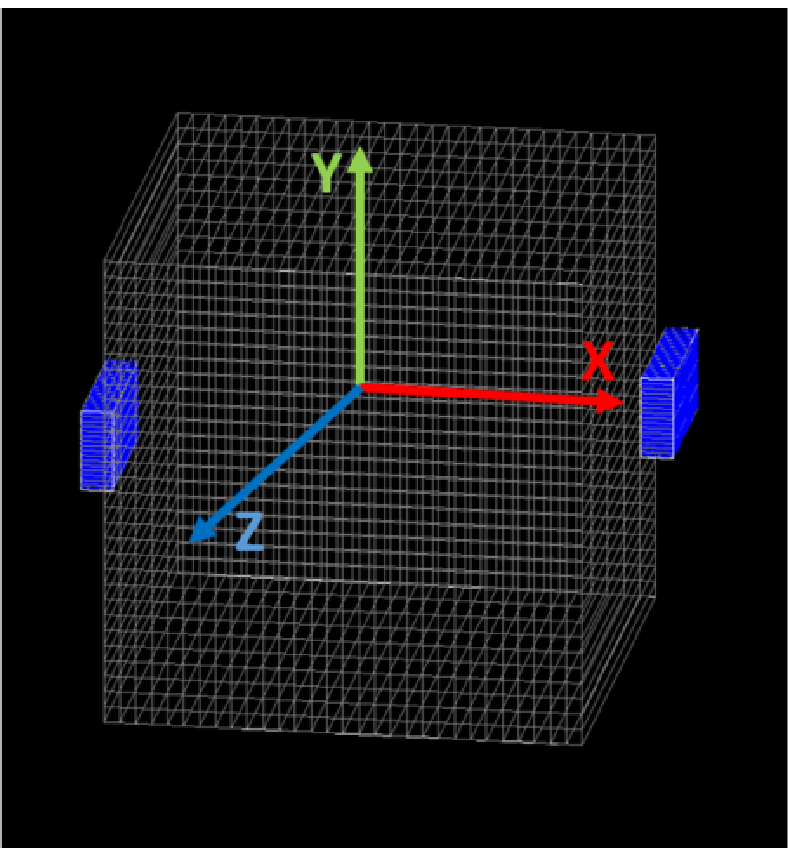}}
\caption{ (a) The simulation setup for the irradiation of the proton beam with the mono-energetic beam and the SOBP case, where the dose of the mono-energetic case is shown in the middle while the dose of the SOBP case is in the lower middle, and the red positions represent the training positions, whereas the blue positions are used for testing. (b) The detector configuration in which the zero position of the detector $z$-axis is aligned with the beam entry face of the phantom, and the detector is translated along the $y$-axis at each position so that the beam is always centered on the detector.}
\label{}
\end{figure}

We used the Geant4-based simulation framework GATE \citep{refe26,refe27} to generate Monte Carlo simulation data for a proton therapy setup using a mono-energetic proton beam and spread out Bragg peak (SOBP) situation separately.   Irradiation of a pencil beam source on an abdominal CT phantom was used for data generation. The phantom dimensions are 302 mm$\times$302 mm$\times$180.5 mm with the spatial resolution of 0.589 mm$\times$0.589 mm$\times$ 0.500 mm. 
The HU range of the phantom was divided into 24 bins according to the scheme described in \citep{refe31}, and the details are summarized in Table~\ref{table:CTcomposition}. 
The beam propagates along the $z$ axis, and the beam profile was assumed to be a Gaussian distribution with standard deviations $\sigma_{x}$ and $\sigma_{y}$ in the iso-center position.
For mono-energetic proton beam, we have performed 900 simulations for five different spot sizes with standard deviations ranging from 3 to 5 mm with an interval of 0.5 mm in both $x$ and $y$ dimensions.   A total of 150 million protons were used as primary particles having a Gaussian energy spectrum with a standard deviation of 1\% from the target proton energy.  The simulations were performed for 10 different positions on the CT phantom along the $y$ axis with a mono-energetic proton beam with energies ranging from 50 to 122 MeV in 4 MeV intervals.  Each irradiation position is spatially 2 cm apart from the other. A visual description of the irradiation configuration on the DICOM phantom is shown in Fig.~\ref{fig:Dicom_dose}. 
The distance between the beam nozzle and the face of the phantom is kept 50 cm. 
The physics list QSP\_BIC\_HP\_EMZ was employed in our study as it showed more accurate results over QSP\_BIC\_HP\_EMY \citep{refe28}. $^{10}$C, $^{11}$C, $^{13}$N, and $^{15}$O have large production cross-sections. Among them, $^{11}$C and $^{15}$O have relatively long lifetimes and larger production cross-sections, which are therefore the focus of our study. The depth dose distribution and the production information of positron-emitting isotopes $^{11}$C and $^{15}$O were stored in a voxelized image using the dose and production-stopping actor available in GATE. Dose and positron activity have been stored with a spatial resolution of  0.589 mm$\times$0.352 mm$\times$0.205 mm. 

For SOBP simulations, a dataset consists of 360 cases with the SOBP of approximately one-quarter of the total range R$_0$ and the maximum beam energy E$_0$  simulated using weighted superposition of mono-energetic proton beams. We selected 12 maximum energies E$_0$s, ranging from 70 to 118 MeV with an energy gap of 4 MeV and irradiated at 10 positions, as shown in Fig. \ref{fig:Dicom_dose}. Three spot sizes with  standard deviations varying from 3 to 4 mm and an interval of 0.5 mm in both $x$ and $y$ dimensions were used. Each energy E$_0$ is divided into 15 weighted fractional energies due to a constraint on computing resources. The simulation scheme and  weights required for various initial energies are calculated using the analytical framework described in \cite{bortfeld1996analytical,jette2011creating}.
\begin{table}[ht]
\caption{Relative elemental composition and density of different materials in accordance with the range of Hounsfield units (HU) \citep{refe31}.} % title of Table
\centering % used for centering 
 \begin{adjustbox}{width=\textwidth}
\begin{tabular}{l c c c c c c c c c c c c c } % centered columns (14 columns)
\hline\hline %inserts double horizontal lines
HU Range& $(\rm g\,cm^{-3})$& H & C & N & O & Na & Mg & P & S & Cl & K & Ca & Ar \\[.5ex] % inserts table
%heading
\hline % inserts single horizontal line
[-1050,-950]&0.001290 & 0    &   0  &  75.6   & 23.2  &     &     &      &     &     &     &  &1.3  \\\
[-950,-120]&0.880110 & 10.3 & 10.5 &   3.1   & 74.9  & 0.2 &     & 0.2  & 0.3 & 0.3 & 0.2 &       \\\
[-120,-82]&0.926691 & 11.6 & 68.1 &   0.2   & 19.8  & 0.1 &     &      & 0.1 & 0.1 &             \\\
[-82,-52]&0.957382 & 11.3 & 56.7 &   0.9   & 30.8  & 0.1 &     &      & 0.1 & 0.1 &     &       \\\
[-52,-22]&0.984227 & 11.0 & 45.8 &   1.5   & 41.1  & 0.1 &     & 0.1  & 0.2 & 0.2 &     &       \\\
[-22,7]&1.011170 & 10.8 & 35.6 &   22    & 50.9  &     &     & 0.1  & 0.2 & 0.2 &     &        \\\
[7,18]&1.029550 & 10.6 & 28.4 &   2.6   & 57.8  &     &     & 0.1  & 0.2 & 0.2 & 0.1 &       \\\
[18,80]&1.061601 & 10.3 & 13.4 &   3.0   & 72.3  & 0.2 &     & 0.2  & 0.2 & 0.2 & 0.2 &       \\\
[80,120]&1.119903 & 9.4  & 20.7 &   6.2   & 62.2  & 0.6 &     &      & 0.6 & 0.3 &     &       \\\
[120,200]&1.111150 & 9.5  & 45.5 &   2.5   & 35.5  & 0.1 &     & 2.1  & 0.1 & 0.1 & 0.1 & 4.5   \\\
[200,300]&1.164774 & 8.9  & 42.3 &   2.7   & 36.3  & 0.1 &     & 3.0  & 0.1 & 0.1 & 0.1 & 6.4   \\\
[300,400]&1.123741 & 8.2  & 39.1 &   2.9   & 37.2  & 0.1 &     & 3.9  & 0.1 & 0.1 & 0.1 & 8.3   \\\
[400,500]&1.282950 & 7.6  & 36.1 &   3.0   & 38.0  & 0.1 & 0.1 & 4.7  & 0.2 & 0.1 &     & 10.1  \\\
[500,600]&1.342190 & 7.1  & 33.5 &   3.2   & 38.7  & 0.1 & 0.1 & 5.4  & 0.2 &     &     & 11.7  \\\
[600,700]&1.401421 & 6.6  & 31.0 &   3.3   & 39.4  & 0.1 & 0.1 & 6.1  & 0.2 &     &     & 13.2  \\\
[700,800]&1.460663 & 6.1  & 28.7 &   3.5   & 40.0  & 0.1 & 0.1 & 6.7  & 0.2 &     &     & 14.6  \\\
[800,900]&1.519900 & 5.6  & 26.5 &   3.6   & 40.5  & 0.1 & 0.2 & 7.3  & 0.3 &     &     & 15.9  \\\
[900,1000]&1.579142 & 5.2  & 24.6 &   3.7   & 41.1  & 0.1 & 0.2 & 7.8  & 0.3 &     &     & 17.0  \\\
[1000,1100]&1.638380 & 4.9  & 22.7 &   3.8   & 41.6  & 0.1 & 0.2 & 8.3  &     &     &     & 18.1  \\\
[1100,1200]&1.697620 & 4.5  & 21.0 &   3.9   & 42.0  & 0.1 & 0.2 & 8.8  & 0.2 &     &     & 19.2  \\\
[1200,1300]&1.756860 & 4.2  & 19.4 &   4.0   & 42.5  & 0.1 & 0.2 & 9.2  & 0.3 &     &     & 20.0  \\\
[1300,1400]&1.816100 & 3.9  & 17.9 &   4.1   & 42.9  & 0.1 & 0.2 & 9.6  & 0.3 &     &     & 21.1  \\\
[1400,1500]&1.875340 & 3.6  & 16.5 &   4.2   & 43.2  & 0.1 & 0.2 & 10.0 & 0.3 &     &     & 21.9  \\\
[1600,5000]&1.946430 & 3.4  & 15.5 &  4.2 & 43.5   & 0.1 & 0.2 & 10.3 & 0.3   &       &     & 22.5  \\[1ex] % [1ex] adds vertical space
\hline %inserts single line
\end{tabular}
\end{adjustbox}
\label{table:CTcomposition} 
\end{table}

\subsection{Detector simulation setup}\label{subsection:DetSimulation}
\begin{figure}[ht]
\center
\subfigure[]{\label{fig:ProtonFlux_Epectrum_a}\includegraphics[scale=.5]{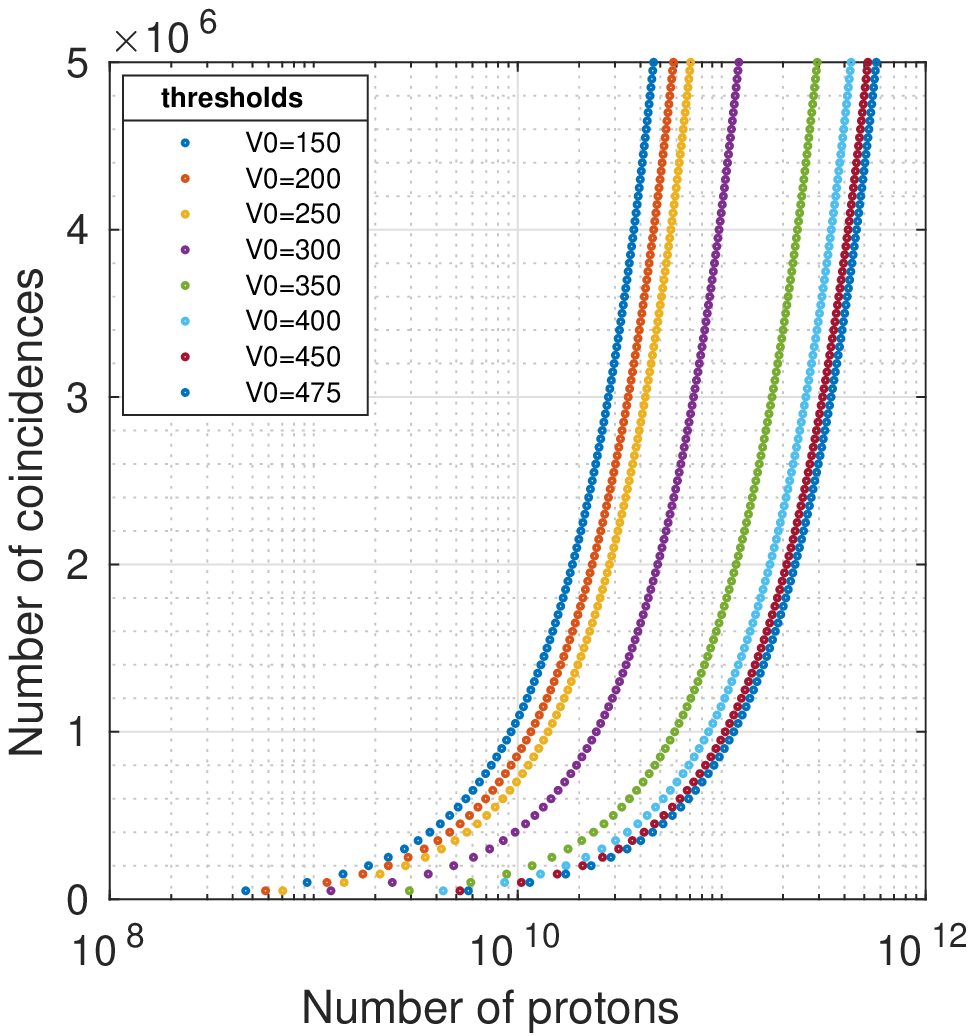}}
\subfigure[]{\label{fig:COIN_energy_a}\includegraphics[scale=.50]{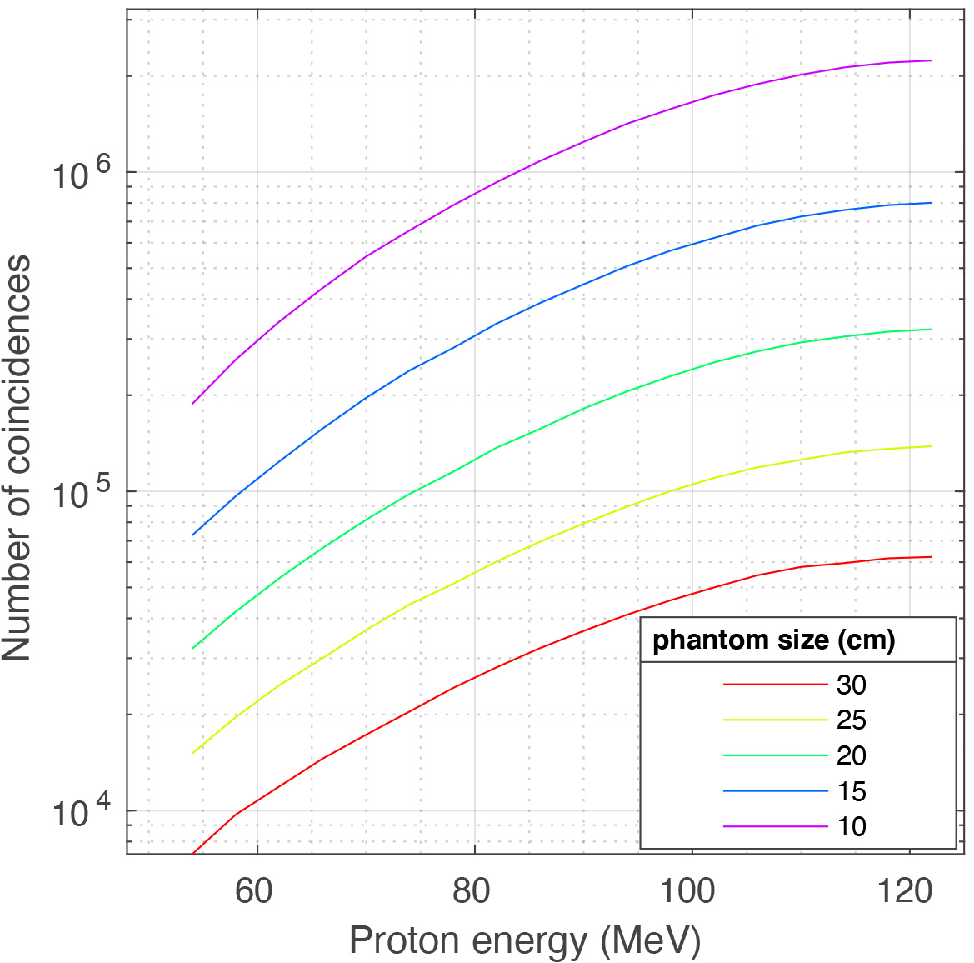}}
\subfigure[]{\label{fig:EnergySpectrumCOIN}\includegraphics[scale=.5]{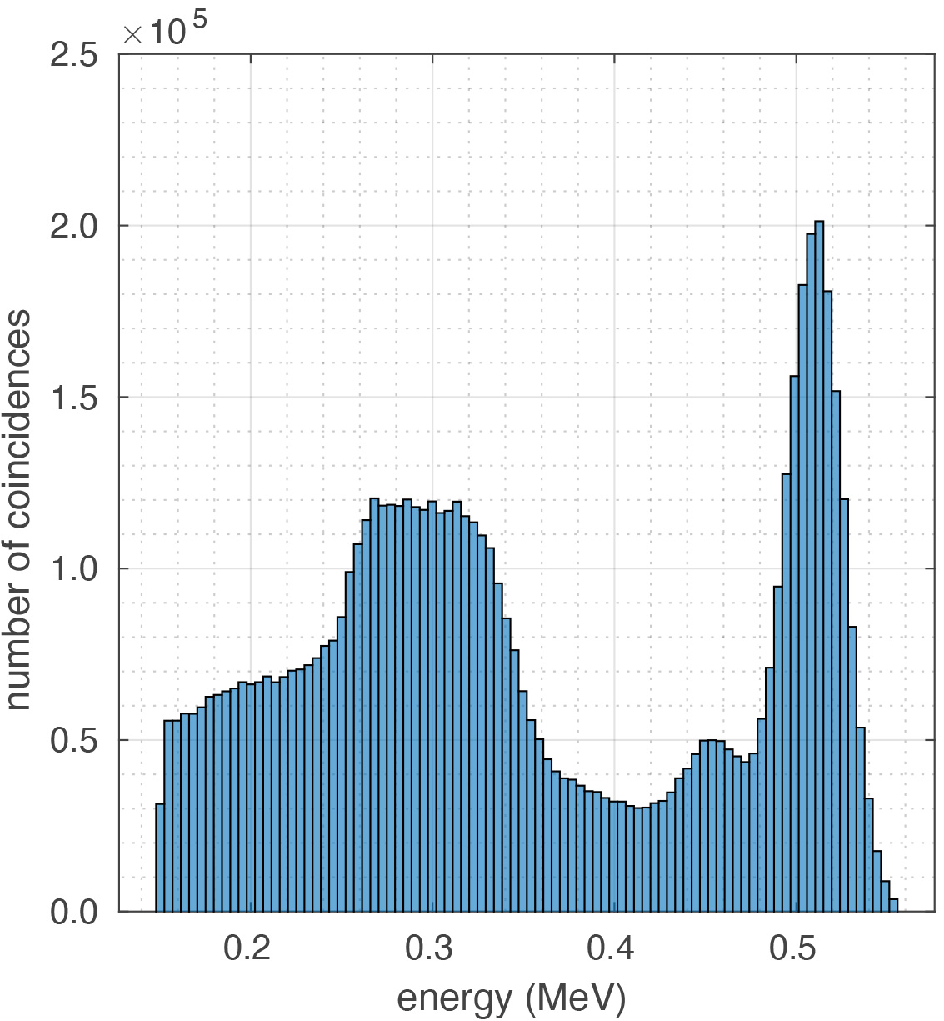}}
\caption{Results of our detector prototype. (a) The relationship between the number of detected coincidences and the number of protons at different detector thresholds for a proton beam energy of 110 MeV. (b) The relationship between the detected coincidence numbers at different proton energies (MeV) and different sizes of human tissue models while varying the detector spacing according to the phantom size, and (c) the simulated energy spectrum of the LYSO crystal.} 
\label{fig:ProtonFlux_Epectrum}
\end{figure}

A palm-sized dual-head PET has 8 modules.   Each module has 64 monolithic LYSO crystals with a size of 3.2 mm$\times$3.2~mm$\times$20~mm, and the spacing between each module is 0.2 mm, resulting in a detector size in $yz$ of 52~$\times$104~mm plane.
The spacing between the dual-head PET modules is maintained at 302 mm apart to accommodate the imaged object between the two flat-panel PET modules. The CT phantom in Fig.\ref{fig:Dicom_dose} is placed in the yz-plane which is parallel to the detector plane as shown in Fig.~\ref{fig:DET_phantom}. Any phantom configuration can be opted based on the clinical requirements. In our study, we have chosen phantom placement  parallel to the detector plane to keep simulation simple. The detector has an energy resolution of 10\% at 511 keV and a temporal resolution of 250 ps FWHM.
%he detector threshold energy for the incoming gamma was kept at 150
The detection threshold is the minimum photon energy accepted by the detector, and we implemented this same idea in our simulation and explored different threshold energy values. The low-energy threshold of the detector was kept at 150 keV to reduce low-energy counts and maintain an acceptable noise level similar to that of a realistic detector. The isotope activities of $^{11}$C and $^{15}$O are imported in the same model configuration that was used during proton beam irradiation as described in Sec.~\ref{subsection:beamSimulation}.
Voxel-based production and stopping information of $^{11}$C and $^{15}$O were used to generate the coincidence data in the palm-sized PET setting.
As we imported the number of isotopes as an activity into GATE, the decay rate constant $\lambda_{^{11}\textrm{C}}$ and $\lambda_{^{15}\textrm{O}}$ of the corresponding isotope were considered to model $^{11}$C and $^{15}$O decay at in-beam PET environment.
The positions of the isotopes can be obtained at the level of one voxel, and finer positions are randomly simulated within each 0.2 mm voxel.
The simulation of the production activity $A(\mathbf{r},t)$ for $^{11}$C and $^{15}$O preserves the relative spatial distribution of the isotopes. The resulting time distribution approximates a realistic simulation when only the two main isotopes are considered, as seen in Eq.~\eqref{eq:activityEquation}. 
\begin{equation} 
    A(\mathbf{r},t)= \lambda_{^{11}\textrm{C}} \times {N}_{^{11}\textrm{C}}(\mathbf{r})\, \textrm{exp}[{-\lambda_{^{11}\textrm{C}}.t}]   +\lambda_{^{15}\textrm{O}}\times {N}_{^{15}\textrm{O}}(\mathbf{r}). \textrm{exp}[{-\lambda_{^{15}\textrm{O}}.t}],   
    \label{eq:activityEquation}
\end{equation}
where $N_{^{11}\textrm{C}}$ and $N_{^{15}\textrm{O}}$ are numbers of isotopes of $^{11}$C and $^{15}$O produced during proton irradiation while $\lambda_{^{11}\textrm{C}}$ and $\lambda_{^{15}\textrm{O}}$  are decay constants of the respective isotopes.
The coincidence data were recorded for 300 seconds, during which about 81\% of $^{15}$O and 15\% of $^{11}$C would have decayed.
%and 51\% of  $^{11}$C isotopes. 
To reduce the simulation time and computational cost, we linearly scaled the activity by a factor of 100 to reach proton fluxes of the order $10^{10}$ since the production of the positron emitters is proportional to the proton fluxes.
The proton fluxes required to generate the coincidence data at 110 MeV proton beam energy under different detection conditions, such as low-energy detector thresholds for our palm-sized prototype with the described simulation setup at a distance of 302 mm, are shown in Fig.~\ref{fig:ProtonFlux_Epectrum_a}.   The number of coincidences recorded at a given proton flux of $\sim10^{10}$ varies with the proton energy and also with the variation in the phantom sizes. We call the distance between the two PET detection planes the detector spacing.   If the detector spacing is adjusted to accommodate different phantom sizes with a marginal gap of 1 mm between the phantom and each detector plane, the number of coincidences detected for different proton energies is shown in Fig.~\ref{fig:COIN_energy_a}.
\subsection{Deep learning model}\label{subsection:deepmodel}
Generally, image mapping problems using measured data can be expressed as:
\begin{equation}
    Y=HX,
\end{equation}
where $Y$ is the measurement data, $H$ is the probability matrix, and $X$ is the sought-after image.   The nature of our problem is pairwise image translation, where the  target $X:\{x_{i}\}$ is in a different domain from the source $Y:\{y_{i}\}$ domain. In the context of our study, $X$ is the dose distribution and $Y$ represents the corresponding measured data, which is a coincidence map in the detectors explained in Sec.~\ref{subsection:DataPreparation}.
The cGAN model uses the conditional variant to initiate the training of the model. Conditional constraints can drive the model to converge to the expected results with significant performance stability. For cGAN models, the overall loss function $\textrm{L}_{cGANs}$  can be expressed \citep{Asako,refe23} in terms of the objective function ${U}(D,G)$ as:
%function $U(G,D)$ of cGAN can be expressed as:

\begin{align}
{L}_{cGANs}(D,G) &=\min_{G}~\max_{D} {U}(D,G) \\
\intertext{and}
{L}_{cGANs}(D,G)&=\textrm{E}_{x \sim Pdata(x)}\left[\log_2 D(x|y)\right]+\textrm{E}_{y \sim Pz(y)}\left[\log_2(1 - D(G(z|y)))\right].
\label{eq:CGAN_math}
\end{align}
The first term in Eq.~\eqref{eq:CGAN_math} corresponds to the encoder function \textrm{E} for the discriminator, while the second term corresponds to the encoder function for the generator. $G:\{y,z\}\rightarrow{x}$ is a joint representation of the generator part of the cGAN model showing the mapping of the dose $X:\{x_{i}\}$ from the projection data $Y:\{y_{i}\}$ mixed with the noise vector $Z:\{z_{i}\}$ from the latent space. Meanwhile, $D$ describes the discriminator part of the model, which computes the probability that the sample $x \sim Pdata(x)$ is coming from the training data instead of $G$. In the Eq.~\ref{eq:CGAN_math}, source $Y:\{y_{i}\}$ distribution combined with random noise $z$ is represented as $z \sim Pz(y)$. $G$ strives to tune the parameters to minimize $\log_2(1 - D(G(z|y)))$, while $D$ tunes the parameters to maximize discriminator loss.
Training with adversarial loss improves image quality, but cannot guarantee similarity to the target image.
To improve accuracy, L1 loss-based regularization $R_{L1}$\citep{l1-lossfunction} is employed in the cGAN model.   $\lambda$ is the weight of the regularization.   Post-regularization the final objective function is expressed as:
\begin{equation}
     G^{\star}=\arg\min_{G}\max_{D}~L_{cGANs}(D,G)+\lambda R_{L1}.
\end{equation}
Our deep learning model  consists of a discriminator and a generator as two primary components where the output of the generator and discriminator are both conditioned on input.
\subsubsection{The architecture of the conditional generative adversarial network}\label{subsection:architecture}
This section will describe the implementation of the cGAN model to convert the coincidence map to a dose map.   The model was implemented using the Nividia GPU Generation GeForce RTX-3090 in the Keras deep learning framework.   As described in Sec.~\ref{subsection:deepmodel}, the model has two main working partners, the discriminator and the generator.
The effective discriminator architecture based on the receptive field (RF) consists of a particular classifier named patchGAN implemented as a down-sampling convolution network.   It makes a patch-based prediction to flag the real or fake image.   It selects a typical size $(70\times70)$ pixels and sweeps over the image to calculate the patch-based probability map of real or fake images instead of computing the whole image at once.   Two types of inputs are fed into the discriminator $X:\{x_{i}\}$ real and $\hat{X}:\{\hat{x}_{i}\}$ generated by the generator and the patch-based probability distribution for both images is calculated.  For model optimization, we have applied the binary cross-entropy (BCE) as the loss function, which is further used by the model to estimate the penalties in the back-propagation step. 
\begin{table}[ht]
\caption{A summary of the hyperparameters of the discriminator.} % title of Table
\centering % used for centering table
\begin{tabular}{c c c} % centered columns (14 columns)
\hline\hline %inserts double horizontal lines
No. & Parameter & Value \\ [1ex] % inserts table
%%heading
\hline % inserts single horizontal line
1 & Alpha for leakyReLU             & 0.2     \\\ % inserting body of the table
2 & Learning rate                   & 0.0002  \\\
3 & Loss weight                     & 1       \\\
4 & Activation function             & Sigmoid \\\
5 & Loss function                   & Binary cross entropy \\\
6 & Optimization algorithm          & Adaptive moment estimation (ADAM) \\\
7 & Beta parameter of (ADAM)        & 0.5     \\[1ex]
\hline
\end{tabular}
\label{table:discriminator} 
\end{table}
\begin{figure}[ht]
\begin{center}
\subfigure[]{\label{fig:generator_transformation}\includegraphics[trim={.9cm 2.4cm 1.1cm 2.5cm},clip,scale=.42]{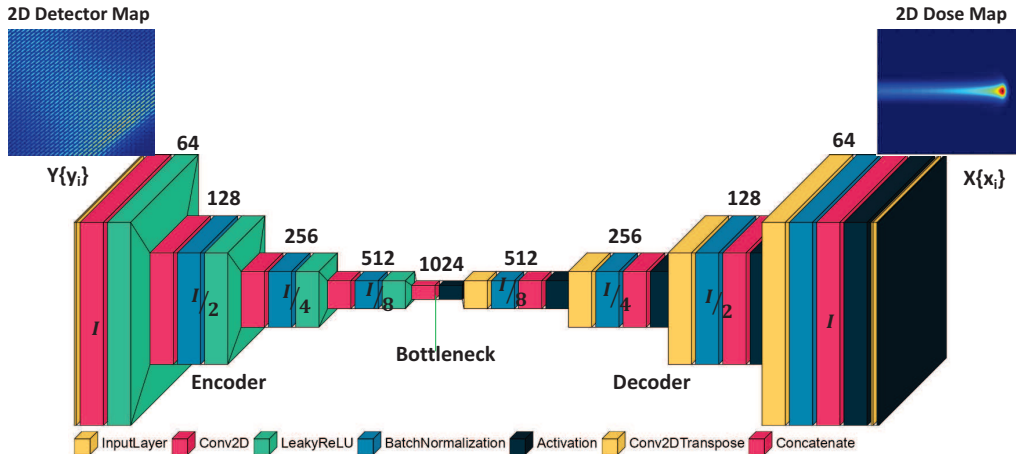}}
\subfigure[]{\label{fig:discriminator_transformation}\includegraphics[trim={6cm 2cm 2.2cm 1cm},clip,scale=.40]{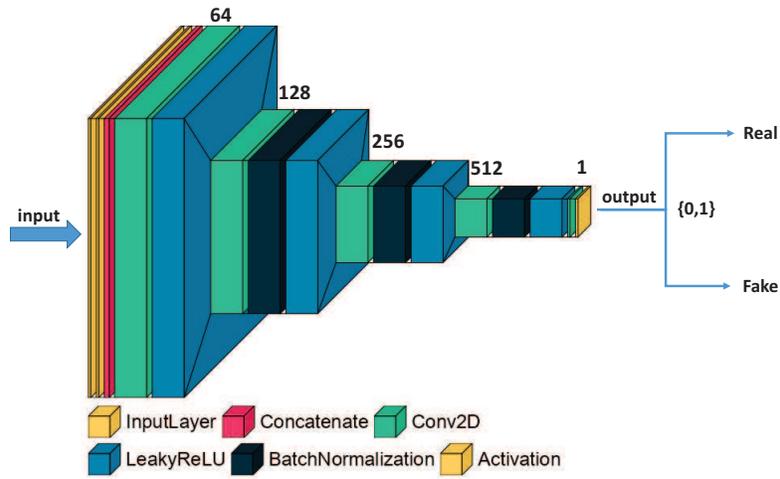}}
%\end{flushleft}
\caption{(a) Block diagrams of the generator taking a coincidence map as input from left and generating the output dose in the $yz$ plane following the axis convention in Fig.~\ref{fig:Dicom_dose} (b) Block diagram of the discriminator taking the generated 2D image from generated as input and giving a binary output.}
\label{fig:cGAN_model_BlockDiagram}
\end{center}
\end{figure}
The generator part of our model has two symmetric encoder and decoder parts.   As shown in Fig.~\ref{fig:generator_transformation}, the encoder part downsamples the input image until the bottleneck, while the decoder part performs upsampling.   Each encoder and decoder section consists of four blocks.   Each normalization block has convolution, pooling, batch normalization, dropout, and activation functions.   Feature information may be lost during downsampling and passed to upsampling using connected layers.   Connections are represented by assigning the same color codes to layers in Fig.~\ref{fig:generator_transformation}.   The upsampled final image has the same size as the input image.   The implementation of the encoder and decoder is performed using unified encoder and decoder functions that, in turn, build a layer block.
The model converges to the final value with scaling of the kernels and layers, but it becomes a computationally expensive task. 

We have used the hyperbolic tangent as the activation function at the end of the downsampling, which predicts the generator output in the range of $[-1,1]$. 
The generator training is performed with two loss functions.   One is the discriminator loss, while the other is $\textrm{L}_{1}$ which describes the mean absolute error (MAE) between the generated and actual distributions.   Since the total loss function of the generator is a linear combination of the discriminator loss and the MAE loss, the weights of the two loss functions are kept at 0.01 and 1.   The higher weight of the MAE loss  enforces the generator to  produce more plausible distributions instead of producing random images in the target domain.   The GAN image translation model encourages keeping the generator stronger than the discriminator.   This goal of convergence to the generated images close to the target images is accomplished by defining a double-component standalone model of the generator and discriminator, where the generator is on top of the discriminator. 

A batch from the real coincidence-dose pair is first used for every training step to update the discriminator.   Then a batch of images generated by the generator is used to update the discriminator, and the discriminator is updated with both real and fake maps.
The generator is fed by the actual 2D coincidence map at the second training phase to predict the dose distribution.   This generated dose map by the generator and the corresponding actual dose map of the corresponding input coincidence map was used to evaluate the loss function on the basis of every single image in the batch and update the generator's weights.
The model is stored after every single epoch for invigilation of the performance.
The optimized values of the hyperparameters for the generator and training sets are given in Table~\ref{table:Discriminator}.

\begin{table}[ht!]
\caption{A summary of hyperparameters for the generator.}
\centering %
\begin{tabular}{c c c}% 
\hline\hline %
No. & Parameter & Value \\ [1ex]%
\hline % 
1 & Alpha for leakyReLU                                & 0.2          \\\ % 
2 & Loss function (L1)                                 & MAE \\\
3 & Loss weight $(w_1)$                                & 1   (BCE)            \\\
4 & Loss weight $(w_2)$                                & 100 (MAE)             \\\
5 & Activation function                                & $\tanh$       \\\
6 & Learning rate                                      & 0.0002       \\\
7 & Batch size                                         & 2    \\\
8 & Epochs                                             & 100  \\\
9 & Optimization algorithm                             & Adaptive moment estimation (ADAM) \\\
10 & Beta parameter of (ADAM)                          & 0.57    \\[1ex]
\hline %inserts single line
\end{tabular}
\label{table:Discriminator} % is used to refer this table in the text
\end{table}
\subsection{Training data preparation scheme}\label{subsection:DataPreparation}
\begin{figure}[ht]
%\begin{frame}
\center
\subfigure[]{\label{fig:CRYSTAL_MAP}{\includegraphics[trim={0cm 3cm 0cm 3cm},clip,scale=.3]{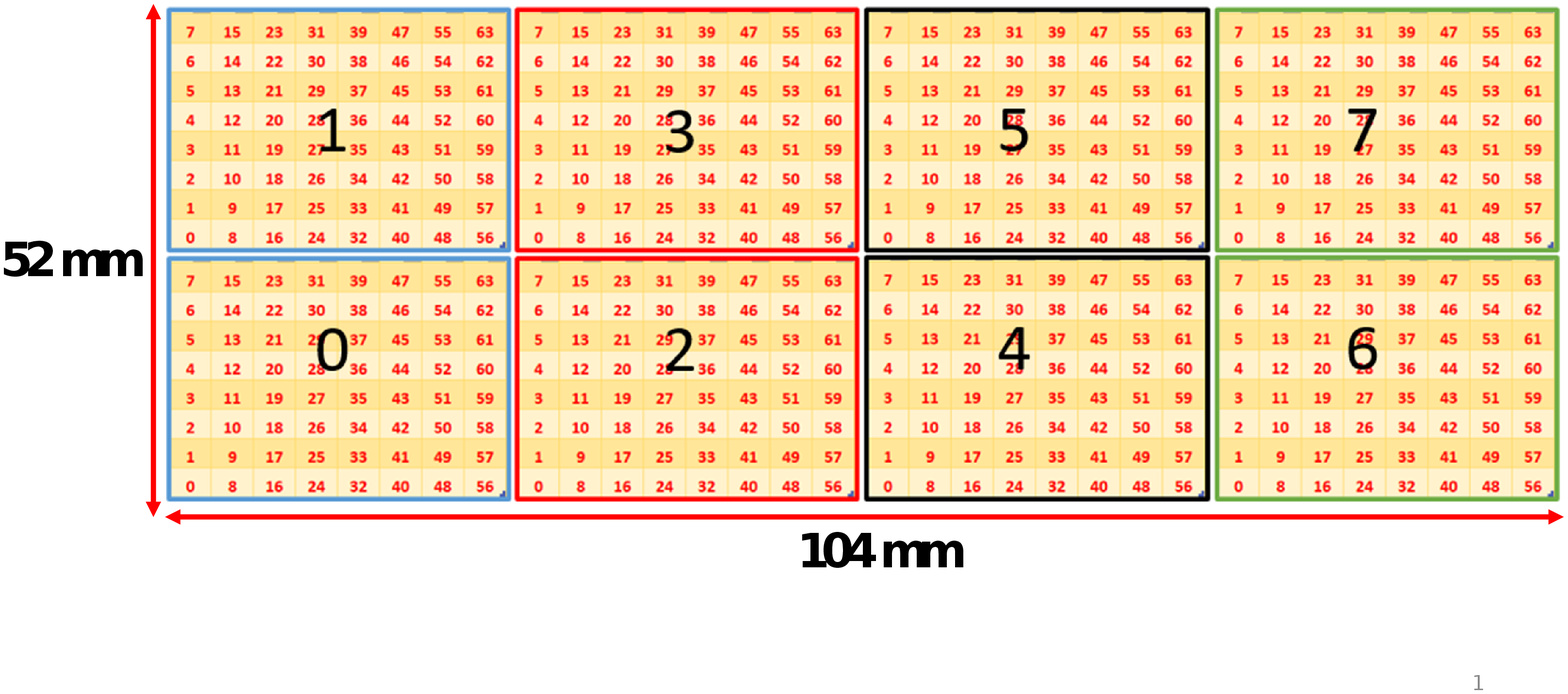}}}  
\subfigure[]{\label{fig:COIN_example110}\includegraphics[trim={0cm 0cm 0cm 0cm},clip,scale=.37]{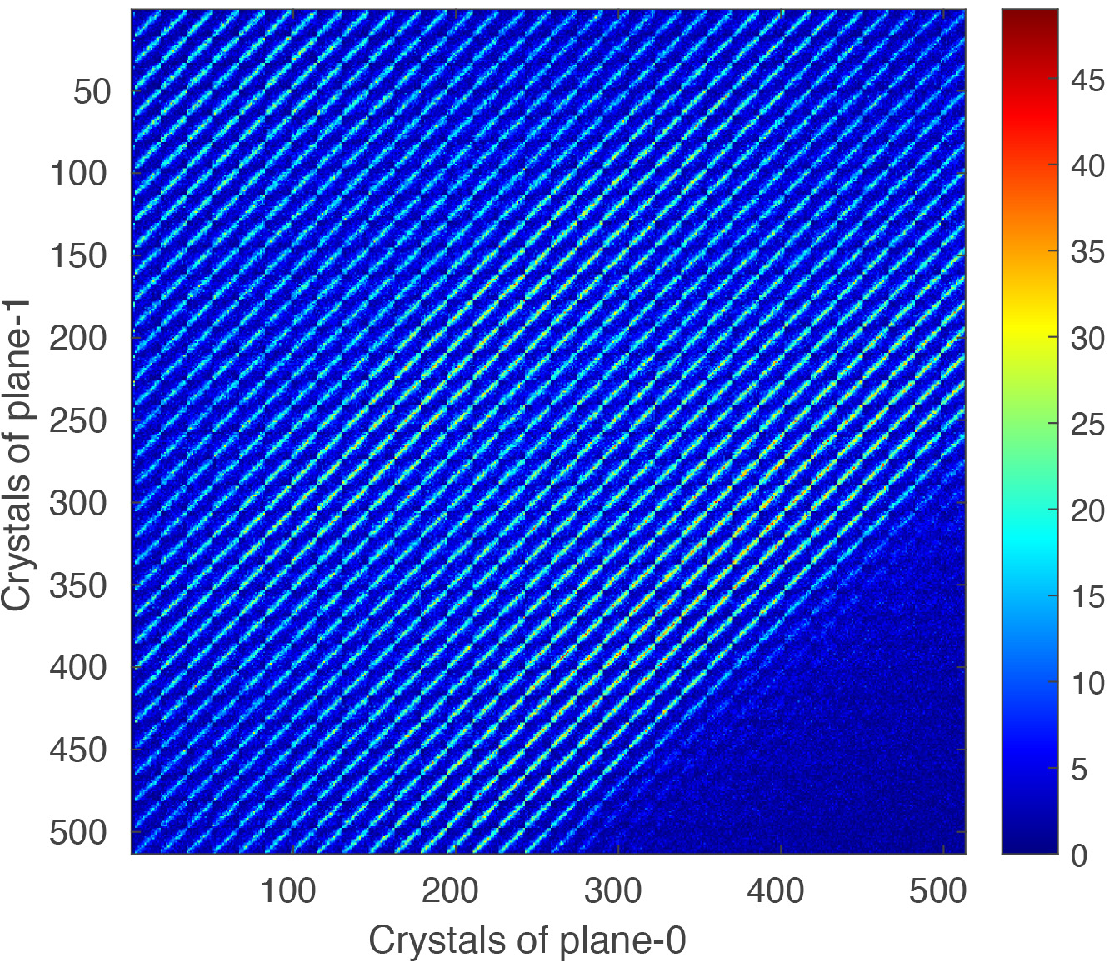}}
\caption{(a) The arrangement of module and crystals in one plane, and (b) a 2D coincidence map for a 110 MeV proton beam as an example.}
\end{figure}

To explore the underlying structure of the data, algorithms based on deep learning models require input chains of training data specific to the target of the problem of interest. The objective of mapping the target of the dose from coincidence information is explained in Sec.~\ref{subsection:deepmodel}. Similar to computer vision problems, dose mapping from the raw coincidence information of the detector also requires a feature representation of the data. The annihilation position of the positron can be estimated by detecting two back-to-back gammas, called a coincidence, using two detector planes. This section describes the organization of the coincidences as a correspondence channel map for two heads of PET, which are further organized according to the photon energy.
 In the detector simulation, the two detector heads of the dual-head PET prototype are tagged as planes 0 and 1.   The geometric layout of each detecting plane is comprised of modules numbered from 0 to 7.   The default arrangement of the crystals in each module is designated from 0 to 63 for plane 0 of the prototype in Fig.~\ref{fig:CRYSTAL_MAP}.
A unique crystal index is generated for each module channel from 0 to 511.   A corresponding map of size $512\times512$ is populated by counting the number of coincidence pairs detected along the column for each crystal in plane 0 and each crystal in plane 1. 
% Line by line means as shown in Fig.~\ref{fig:2D_HEAT_MAP}.
A typical 2D coincidence map is shown in Fig.~\ref{fig:COIN_example110}.   We employed both direct coincidence and cross-coincidence to make our data representation similar to the Michelograms \citep{refe_m} originally used for the reconstruction of 3D PET images.
  
After obtaining all coincidence maps and the corresponding dose maps, we combined them into the paired source and target images, as shown in Fig.~\ref{fig:generator_transformation}.
Two datasets were generated.   The first dataset consists of coincidence maps and the corresponding dose maps as 2D image pairs for different detector thresholds.   For the second dataset, the data containing the coincidence map and the corresponding dose map as 2D image pairs have three channels.
Three-channel 2D coincidence maps were prepared by filtering coincidences in the three energy windows.   The three maps are stacked as three channels of a 2D map.   In the remainder of this paper, the three-channel coincidence map and the corresponding dose-paired dataset are designated as the three-energy window (EW3) dataset.   The energy windows are: a) 150 keV to 250 keV, b) 250 keV to 475 keV and c) $>$474 keV.
We irradiated 10 locations at 18 energy levels using 5 different spot sizes and generated 900 corresponding image data. The seven lower positions marked red in Fig.~\ref{fig:Dicom_dose} are reserved for training, and the three upper positions marked blue along the vertical line are reserved for testing purposes. We split the total data into training data (70\%) and test data (30\%) for model evaluation.   We used 9\% of the training data as validation data for hyper-parameter tuning. The same 70\% training data and 30\% test data scheme is used for 360 SOBP cases, and hyper-parameter values remain the same as those used for training mono-energetic proton beam.
\subsection{Model evaluation metrics}\label{subsection:modelEvaluation}
To evaluate the prediction of Bragg's peak position, we used the shift in the BP values of the predicted profile and the true profile.   The 2D profiles were converted to 1D profiles and the 5\% maximum values around the peak position of the true and predicted profiles were fitted to a Gaussian fit function.
The ground truth values are known in simulation-based studies, allowing learning-based methods to quantify performance evaluations.   We selected three variables as our model evaluation metrics.   These metric variables are the mean relative error (MRE), the fractional absolute dose error ($\delta{D}$), and the shift of the Bragg peak (BP) denoted by ($\delta{R}$). 
We evaluated MRE and $\delta{D}$ at two regions of interest (ROIs).   The first region of interest, called ROI$_1$, consists of the entire beam profile region, while the second region of interest, ROI$_{2}$ contains the region enclosed between the 50\% peak values on the proximal and distal sides of the Bragg peak.   MRE and $\delta{D}$ are relevant variables to grade the intensity of the predicted feature, which is related to the verification of the dose in the final image.   The metrics are defined as: 
\begin{equation}
    \textrm{MRE}=\frac{1}{n}\sum_{i}^{n}\frac{|\hat{s_{i}}-s_{i}|}{\max(s)},
\end{equation}
where $\hat{s_{i}}$ and $s_i$ respectively represent the predicted and true values in a pixel in an ROI and $n$ represents the total number of pixels within. In the MRE formula, $\max(s)$ represents the peak value of the true profile. The fractional absolute dose error is defined as:
\begin{equation}
  \delta{D}=\frac{ \bigg|\int_{-z_{d}}^{+z_{d}} \boldsymbol{\hat{D}}(z)dz  - \int_{-z_{d}}^{+z_{d}} \boldsymbol{D}(z)dz \bigg|}{\int_{-z_{d}}^{+z_{d}} \boldsymbol{D}(z)dz}, 
\end{equation}
where $\boldsymbol{\hat{D}}$ denotes the 1D projection of the predicted dose profile, while $\boldsymbol{{D}}$ denotes the true 1D projection of the dose profile of the simulation used as the ground truth.   Here $-z_{d}$ denotes the position of the dose profile on the left of the BP,  while the $+z_{d}$ denotes the dose position to the right of the BP.   The shift in BP assesses the performance using a geometric feature related to range verification as:
\begin{equation}
   \delta{R}=\hat{R}_{BP}-R_{BP}.
\end{equation}
Here, $\hat{R}_{BP}$ shows the BP position of the predicted profile, $R_{BP}$ shows the BP position of the true profile and $\delta{R}$ is the shift in the predicted profile in mm. 

\section{Results}\label{section:Results}
\subsection{Prediction of dose, and positron-emitting isotope distribution }\label{subsection:DoseMapping}
\begin{figure}[ht]
\begin{multicols}{6}
\subfigure[]{\label{fig:Generated_dose_a}\includegraphics[trim={0cm 0cm 0cm 0cm},clip,scale=.35]{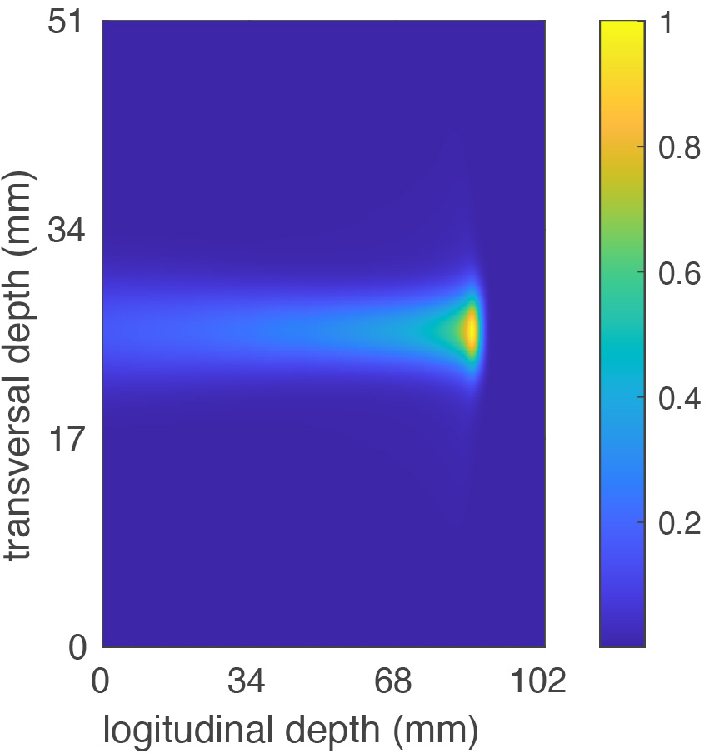}}%\par
\subfigure[]{\label{fig:RealDose_b}\includegraphics[trim={0cm 0cm 0cm 0cm},clip,scale=.35]{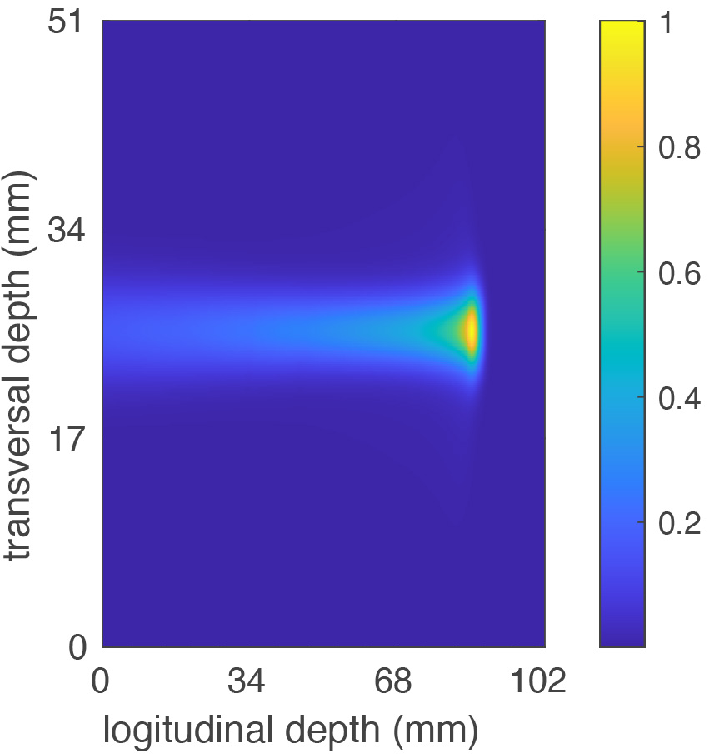}}%\par
~\subfigure[]{\label{fig:DoseDifference_c}\includegraphics[trim={0cm 0cm 0cm 0cm},clip,scale=.35]{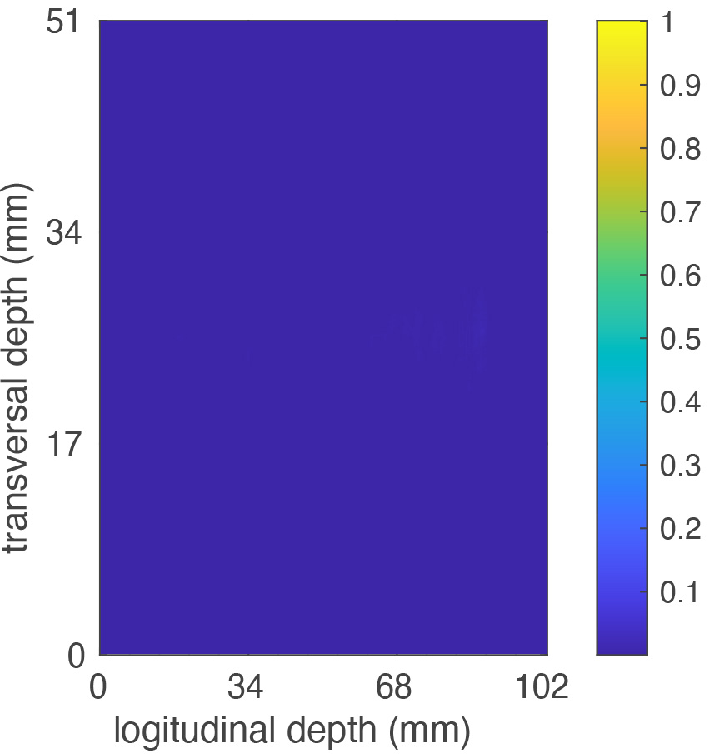}}%\par
%\end{multicols}
%\begin{multicols}{3}
\subfigure[]{\label{fig:GenPAG_a}\includegraphics[trim={0cm 0cm 0cm 0cm},clip,scale=.35]{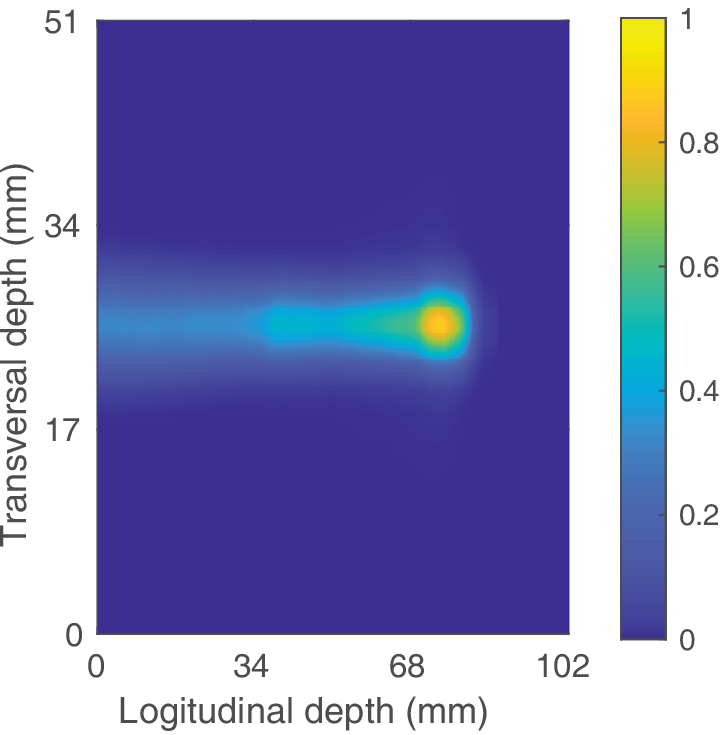}}%\par
~\subfigure[]{\label{fig:RealPAG_b}\includegraphics[trim={0cm 0cm 0cm 0cm},clip,scale=.35]{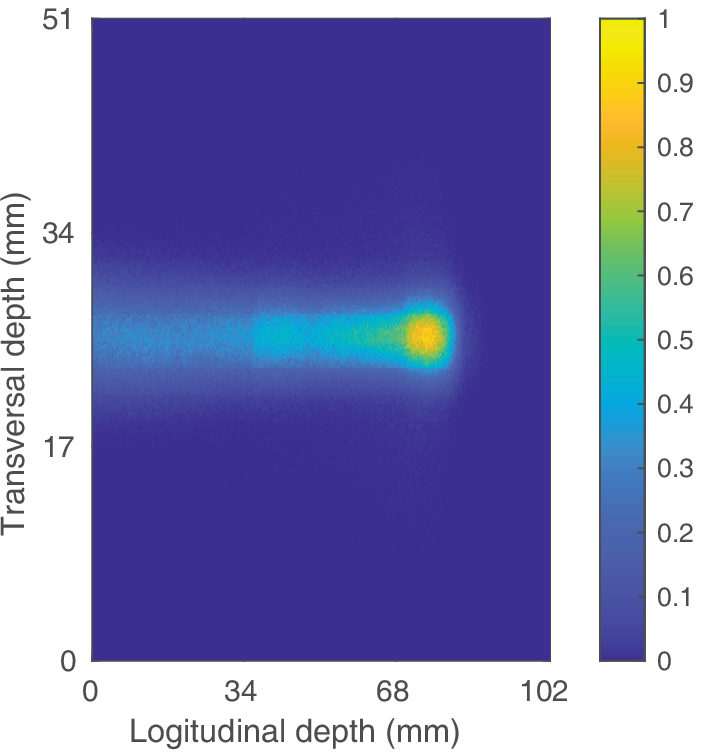}}%\par
~\subfigure[]{\label{fig:PagDiff_c}\includegraphics[trim={0cm 0cm 0cm 0cm},clip,scale=.35]{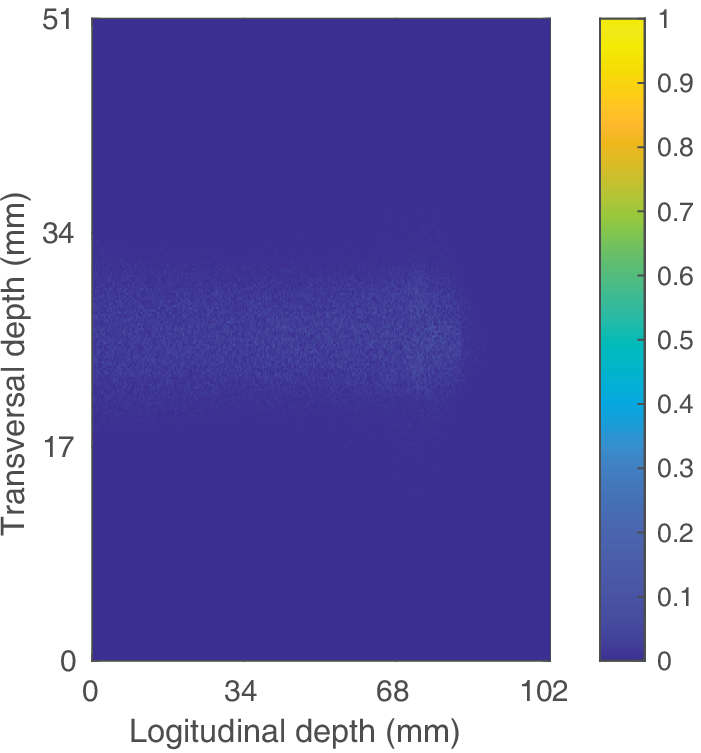}}%\par
%\end{center}
\end{multicols}
\begin{multicols}{5}
\subfigure[]{\label{fig:Material}\includegraphics[trim={0cm 0cm 0cm 0cm},clip,scale=.40]{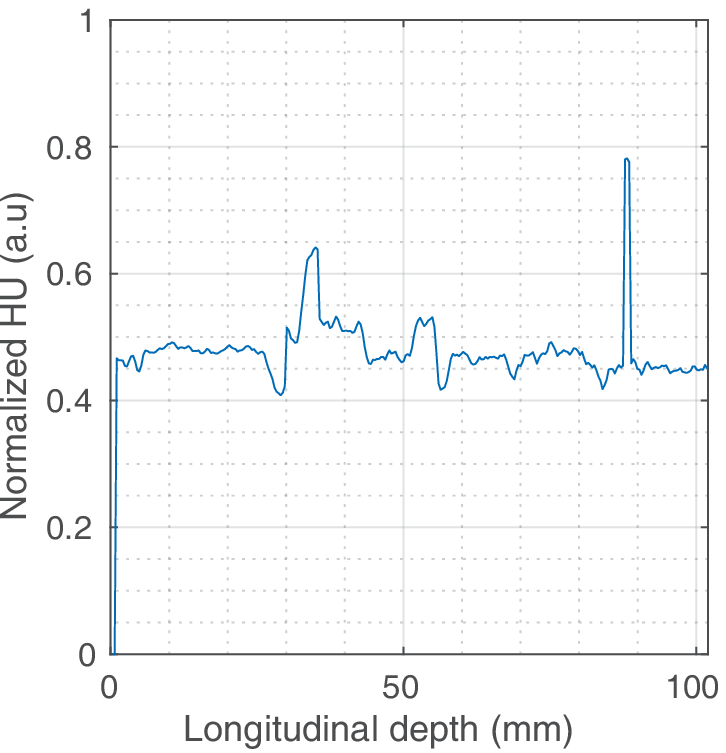}}%\par
\subfigure[]{\label{fig:Dose_1d}\includegraphics[trim={0cm 0cm 0cm 0cm},clip,scale=.35]{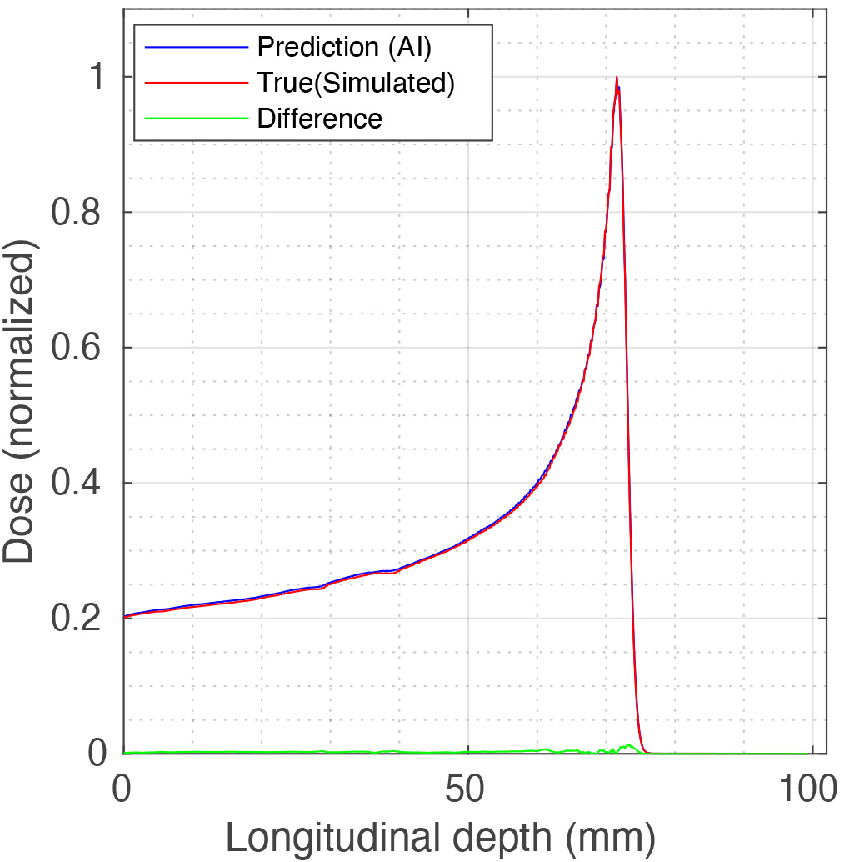}}%\par
\subfigure[]{\label{fig:PAG_1d}\includegraphics[trim={0cm 0cm 0cm 0cm},clip,scale=.3]{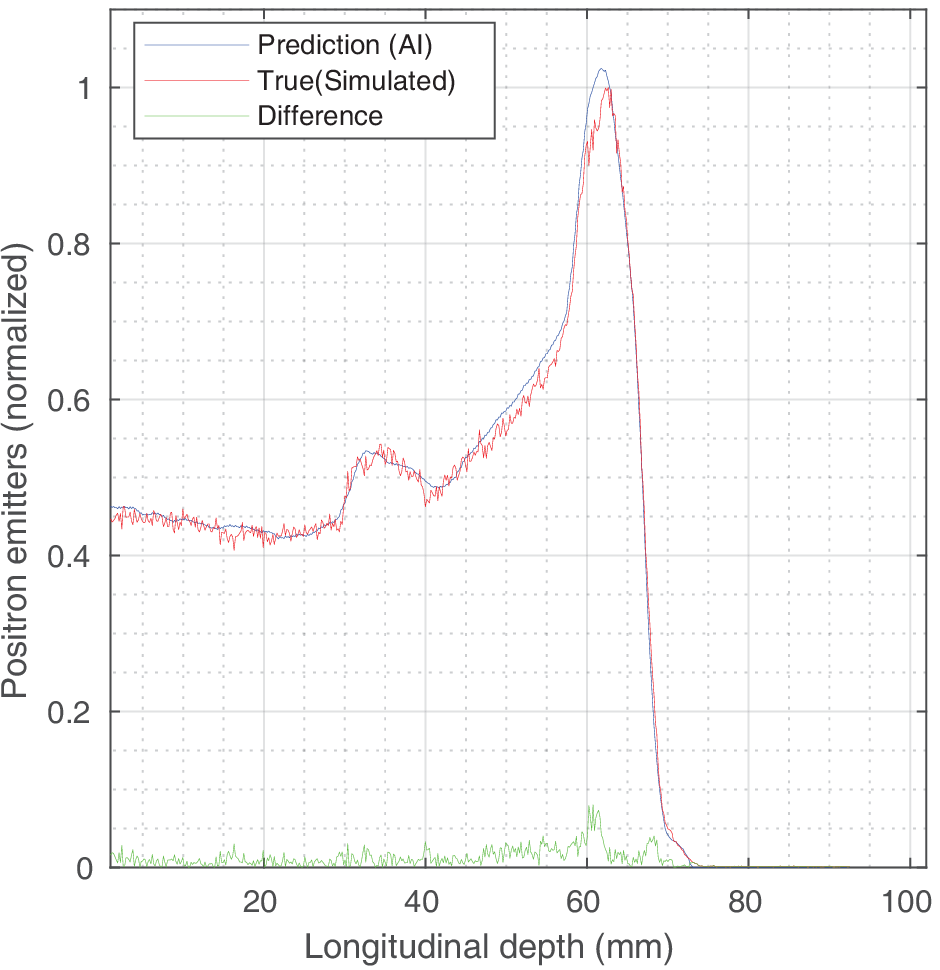}}%\par
\subfigure[]{\label{fig:Dose_1d_trans}\includegraphics[trim={0cm 0cm 0cm 0cm},clip,scale=.3]{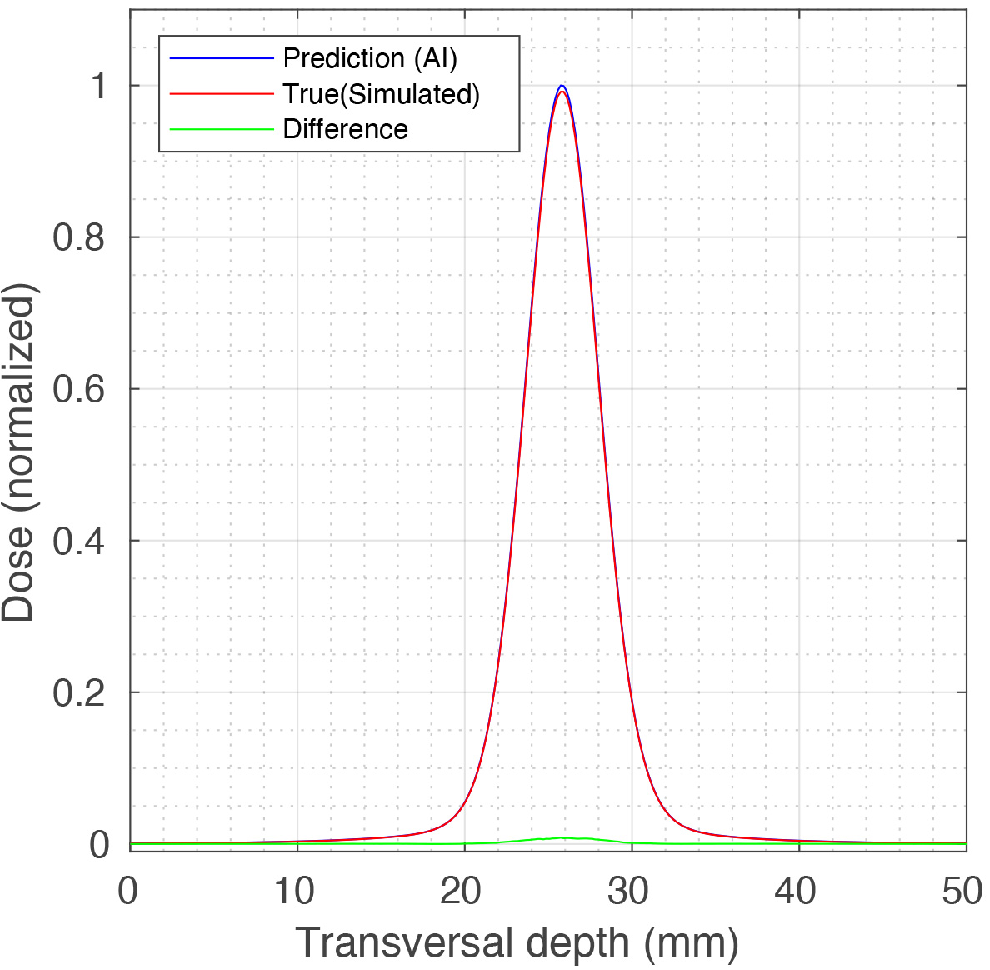}}%\par
\subfigure[]{\label{fig:PAG_1d_trans}\includegraphics[trim={0cm 0cm 0cm 0cm},clip,scale=.3]{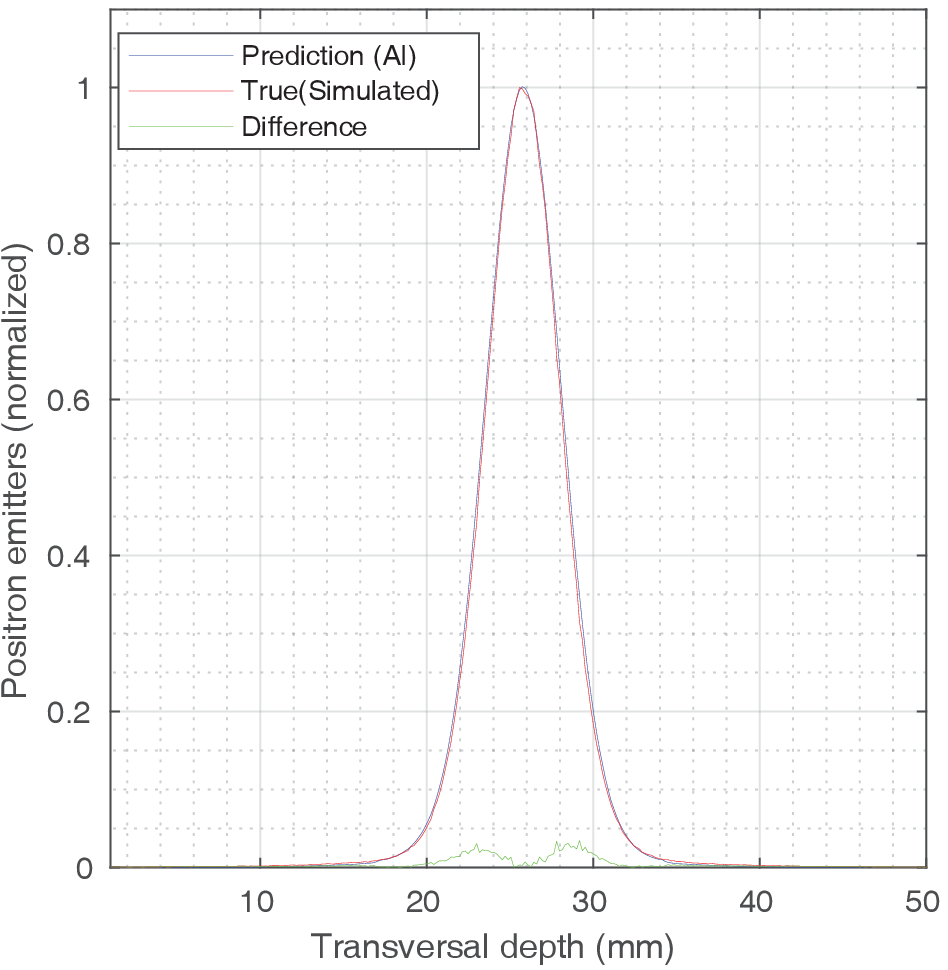}}%\par
%\end{center}
\end{multicols}
\caption{Results for a 110 MeV monoenergetic proton beam. (a) The dose distribution generated using the AI model, (b) the true simulated dose distribution, and (c) the absolute difference between the generated and true dose distribution. ( d) The $\beta^+$ profile generated using the AI model, (e) the real simulated profile of $\beta^+$,  and (f) the absolute difference between the generated and real $\beta^+$ distributions. (g) The material density along the irradiation path measured in arbitrary units (a.u.) by normalization to the peak of Hounsfield units, (h) the 1D dose normalized to the target peak position along the irradiation direction, (i) the 1D depth distribution of $\beta^+$ activity, normalized to the target peak position along the irradiation direction, (j) the 1D depth dose in the transverse direction, and (k) the 1D positron-emitting isotopes profile in the transverse direction. The blue, red, and green curves represent the AI prediction, the true contour, and the difference between the predicted and true values, respectively.}
\label{fig:generalCapbility}
\end{figure}

\begin{figure}[ht]
\begin{center}
\subfigure[]{\label{fig:RME_Histogram}\includegraphics[scale=.4]{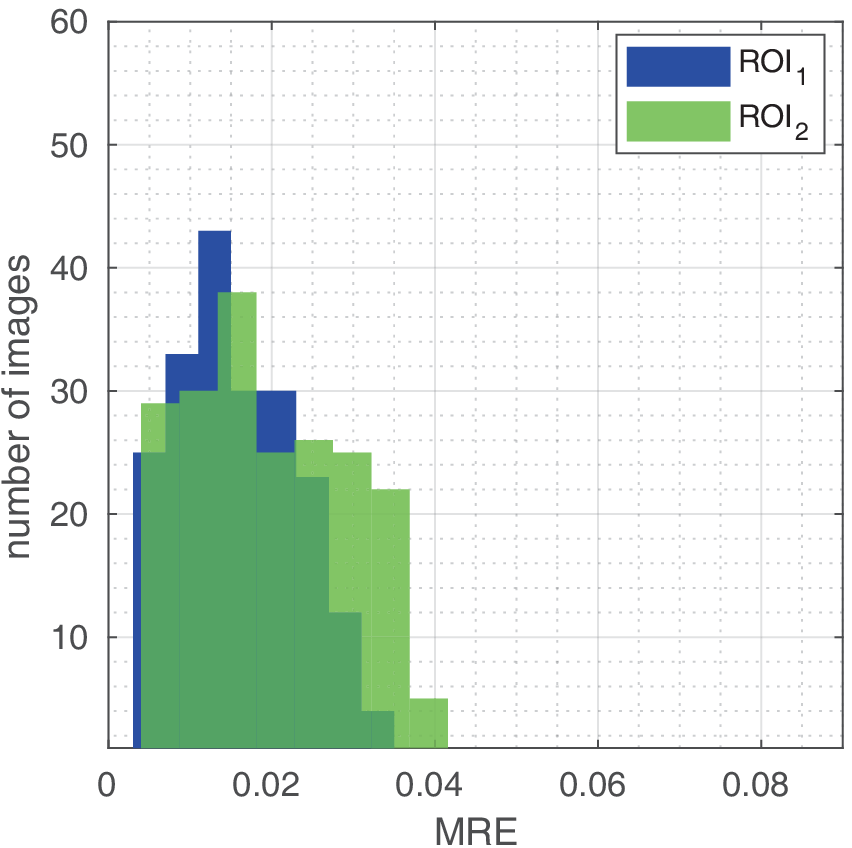}}
~\subfigure[]{\label{fig:AbsDose_Histogram}\includegraphics[scale=.4]{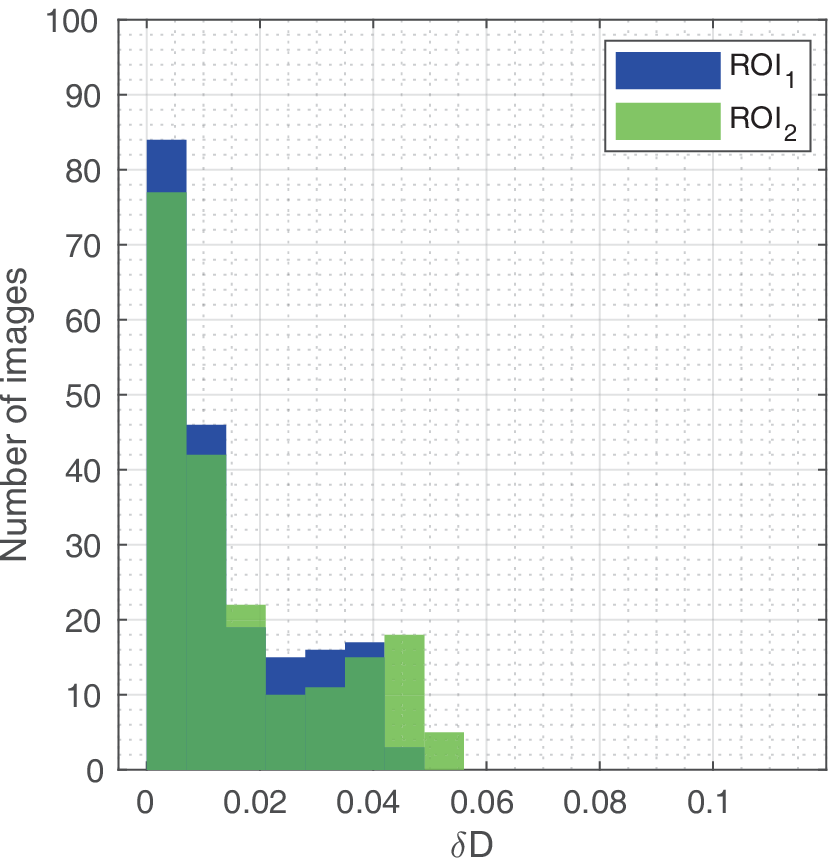}}
~\subfigure[]{\label{fig:ShiftInBP_HistTestData}\includegraphics[scale=.4]{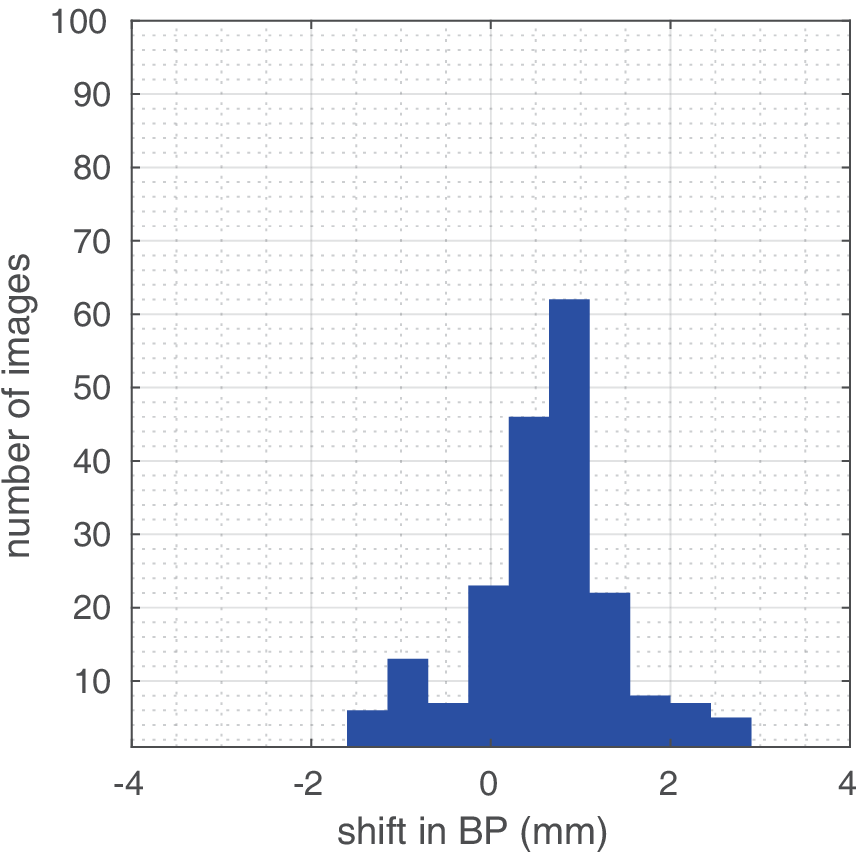}}
\end{center}
\caption{ (a) MRE histogram of the test dataset, (b) histogram of the absolute dose difference ($\delta{D}$) of the entire test dataset in ROI$_{1}$ and ROI$_{2}$, (c) BP ($\delta{R}$) shift histogram for the full test dataset.}
\label{fig:DoseRangeHistTest}
\end{figure}

 The results of dose prediction and mapping of positron-emitting isotopes are shown in this section.   The key is to consider that the results of dose mapping and  positron-emitting isotope mapping correspond to two  independently trained models, one for dose prediction and the other for isotope distribution prediction.   The first part of this section shows the general predictive capability of the dose.   Fig.~\ref{fig:generalCapbility}(a)-(c) shows the predicted dose map, the true dose map, and the absolute difference between the prediction and true values at 110 MeV energy in the test dataset.   In each plot, the horizontal axis represents the longitudinal depth ($z-$axis) along the beam irradiation direction, and the vertical axis represents the transverse depth ($y-$axis) perpendicular to the beam direction.   The predicted profile in Fig.~\ref{fig:generalCapbility}(a) agrees well with the ground truth in Fig.~\ref{fig:generalCapbility}(b).   To gain more insight, the values of the pixels along the $y-$ axis are summed to obtain the 1D dose distribution as a function of longitudinal depth.   Fig.~\ref{fig:Dose_1d} shows that the discrepancy between the true and predicted BP locations is about 0.35 mm, less than 1\% of the BP peak position, and an absolute dose uncertainty of less than 2\%.   Similarly, pixel values along the longitudinal depth are summed to obtain the transverse profile of the dose in Fig.~\ref{fig:Dose_1d_trans}.  
The general prediction capability of the positron-emitting isotopes can be observed in Fig.~\ref{fig:generalCapbility}(d)-(f).   The integrated depth profile of the positron-emitting isotopes of $^{11}$C and $^{15}$O was predicted using an AI-based model that took detector coincidence data as input for prediction.   The image size remains the same as the dimension of the detector plane.   The absolute difference in positron activity is predicted to be within the uncertainty of $\pm{2\%}$, while the change in BP was predicted to be $\pm{0.5}$ mm, which is less than ${1.3\%}$ of the longitudinal depth. 1D profiles of positron emitters can be seen in Fig.~\ref{fig:PAG_1d} and Fig.~\ref{fig:PAG_1d_trans}.
This section shows the evaluation of the model on the test data using the mean of the post-training metric variables MRE, $\delta{D}$, and $\delta{R}$.   MRE, fractional dose shift, and fractional shift in BP for complete test data were calculated from the histograms in Fig.~\ref{fig:DoseRangeHistTest}(a)-(c).   The results with the corresponding error bars are described in Table~ \ref{table:TestDataSummary}.   The results are then presented within 95\% confidence intervals.

\begin{figure}[ht]
\begin{center}
\subfigure[]{\label{fig:SOBP_gen}\includegraphics[scale=.4]{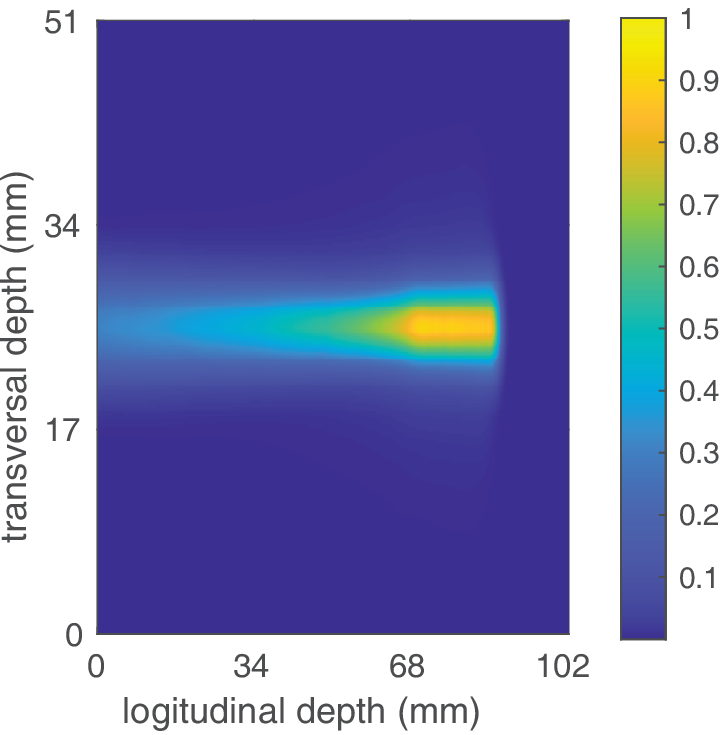}}
~\subfigure[]{\label{fig:SOBP_True}\includegraphics[scale=.4]{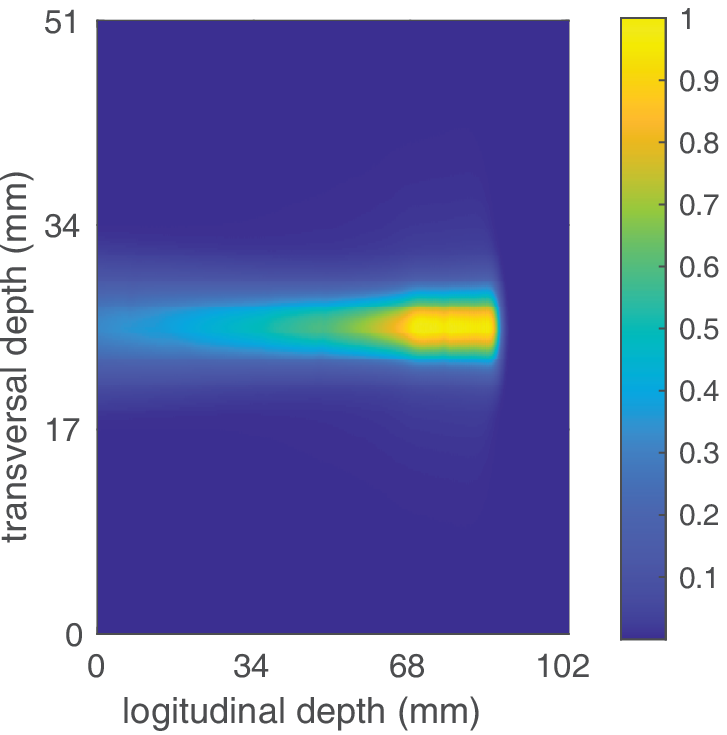}}
~\subfigure[]{\label{fig:SOBP_DIFF}\includegraphics[scale=.4]{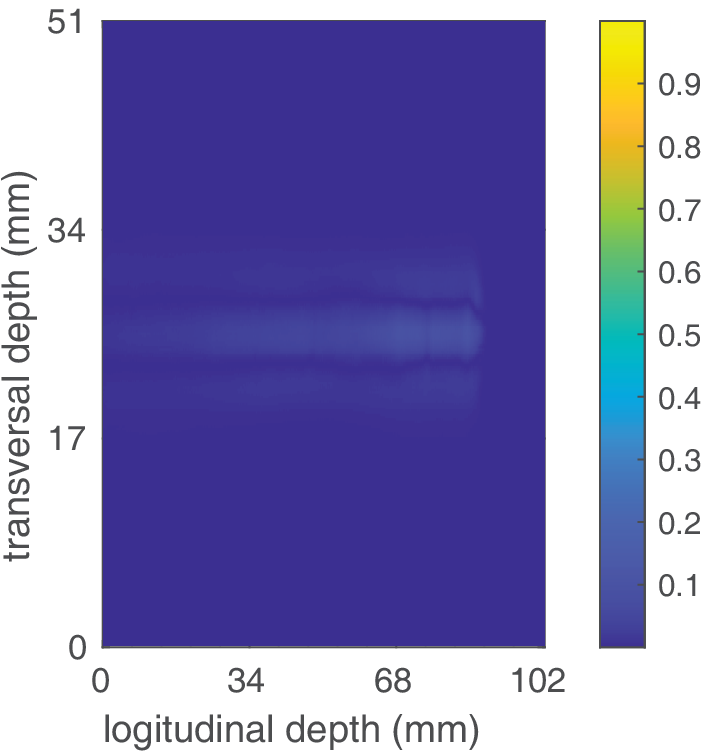}}\\
\subfigure[]{\label{fig:Material_Sobp}\includegraphics[trim={0cm 0cm 0cm 0cm},clip,scale=.435]{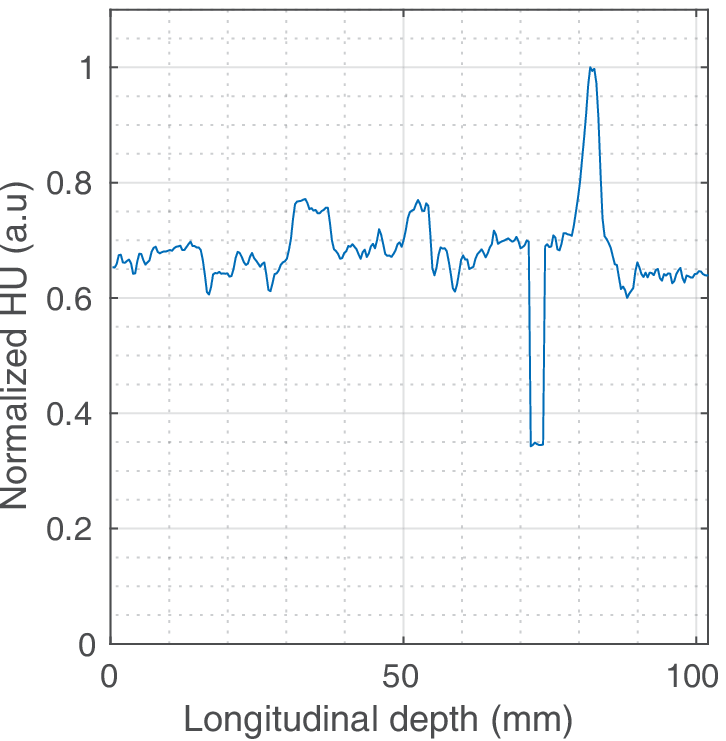}}%\par
~\subfigure[]{\label{fig:SOBP_True_1D_along}\includegraphics[scale=.3]{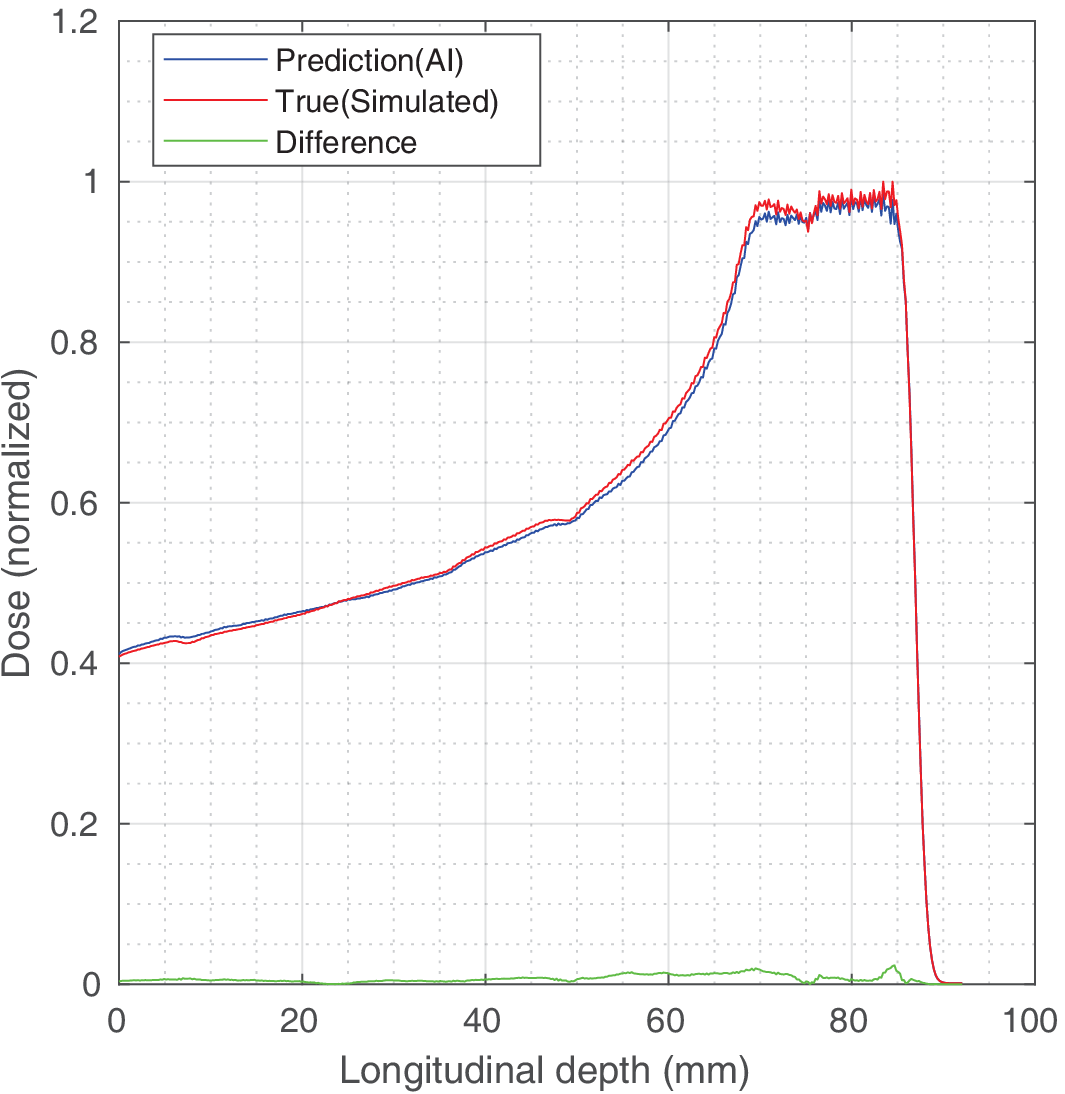}}
~\subfigure[]{\label{fig:SOBP_True_1D_trans}\includegraphics[scale=.3]{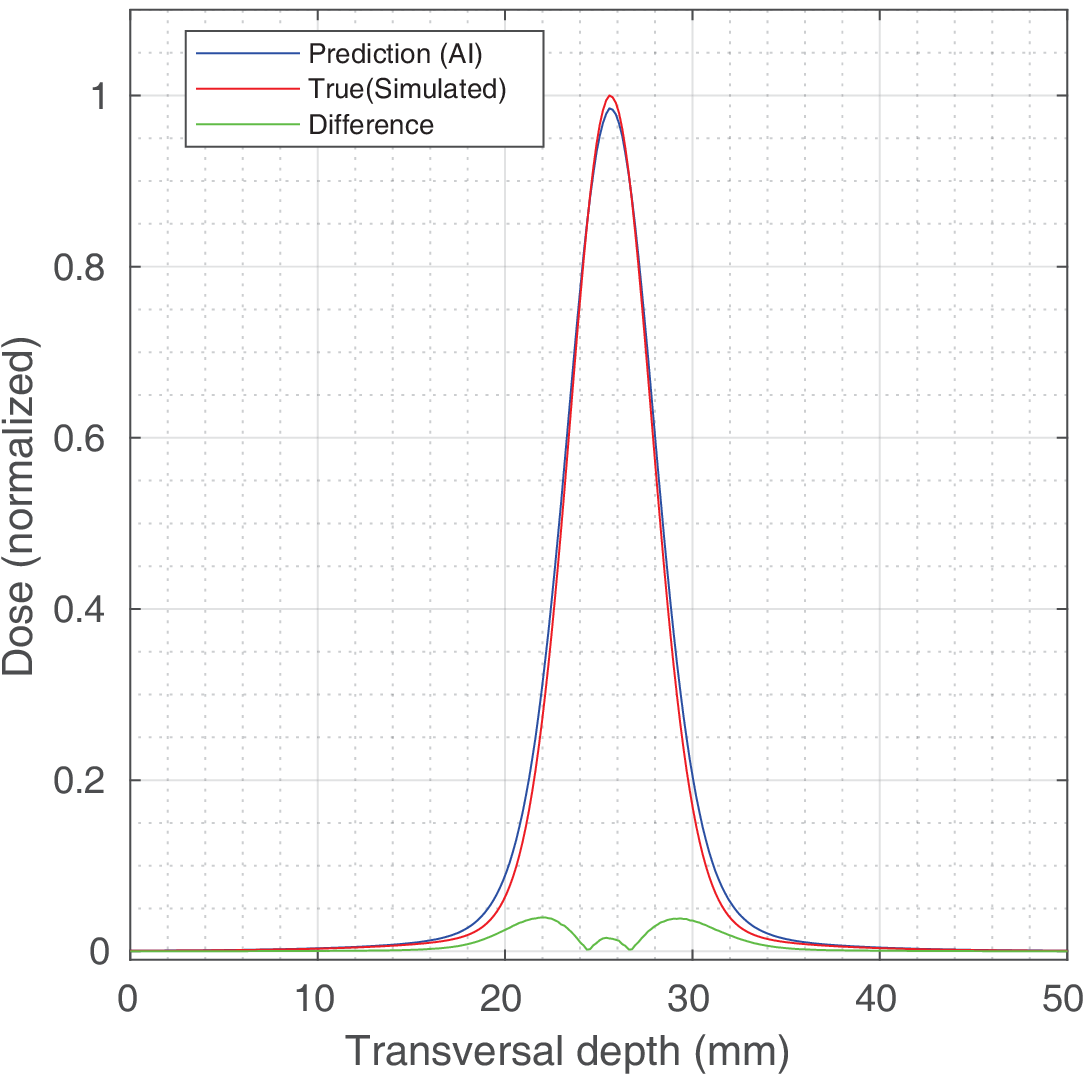}}
\end{center}
\caption{SOBP results for the dose at a maximum proton beam energy of 114 MeV. (a) The predicted SOBP profile using the AI model, (b) the true simulated SOBP dose profile, and (c) the absolute difference in the generated and true dose profiles. (d) The material density of the irradiation path normalized to the peak of Hounsfield units presented in arbitrary units (a.u), (e) 1D dose profile normalized to the peak position of the target along the irradiation direction, and (f) 1D depth dose in the transverse direction.}
\label{fig:SOBP_profile}
\end{figure}

The results of SOBP are shown in Fig.~\ref{fig:SOBP_profile} for a proton beam with a maximum energy of 114 MeV from the test data. Fig.~\ref{fig:SOBP_profile}(a)-(c) shows the predicted dose map, the true dose map, and the absolute difference between the prediction and true values at 114 MeV energy in the test dataset. The dose difference between real and predicted is less than 2.5\% and a shift in BP is less than 1 mm can be observed in the 1D profile shown in Fig.~\ref{fig:SOBP_True_1D_along} along the longitudinal depth. The 1D dose profile along the transverse depth is shown in Fig.~\ref{fig:SOBP_True_1D_trans}. The mean values of MRE and $\delta{D}$  for SOBP in ROI$_1$ and ROI$_2$ are described in Table~\ref{table:TestDataSummary} under the SOBP title. ROI$_2$ for SOBP cases is comprised of the spread out Bragg peak region only.

\subsection{Performance evaluation}\label{subsection:dataAna}
\begin{figure}[ht]
\begin{center}
\subfigure[]{\label{fig:MRE_epoch_b}\includegraphics[scale=.4]{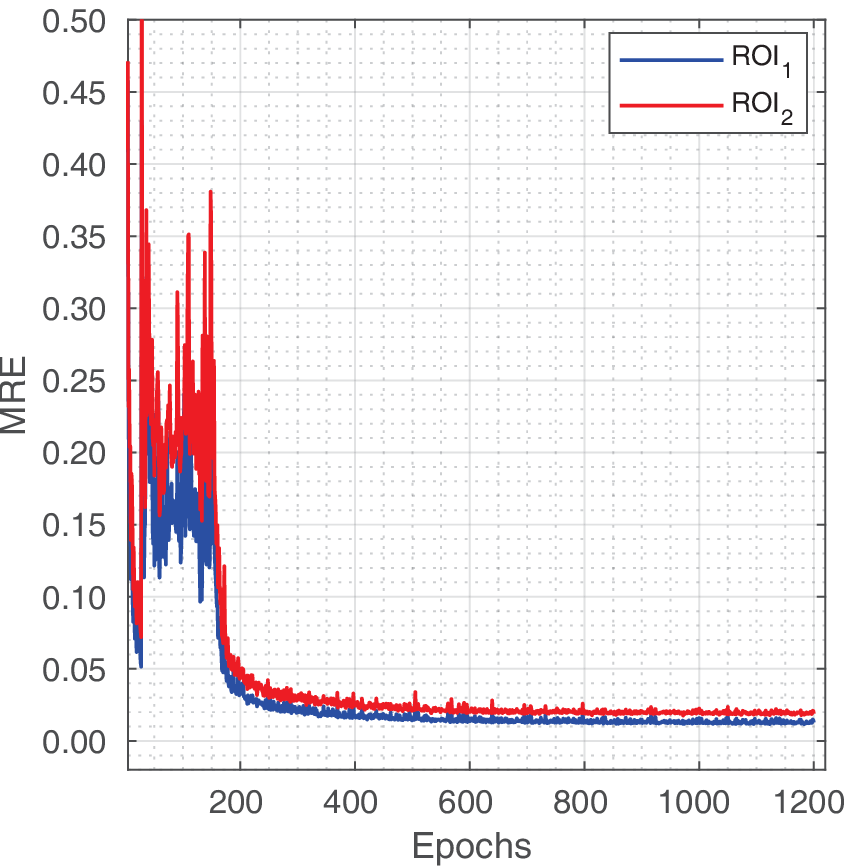}}
~\subfigure[]{\label{fig:FracDose_c}\includegraphics[scale=.4]{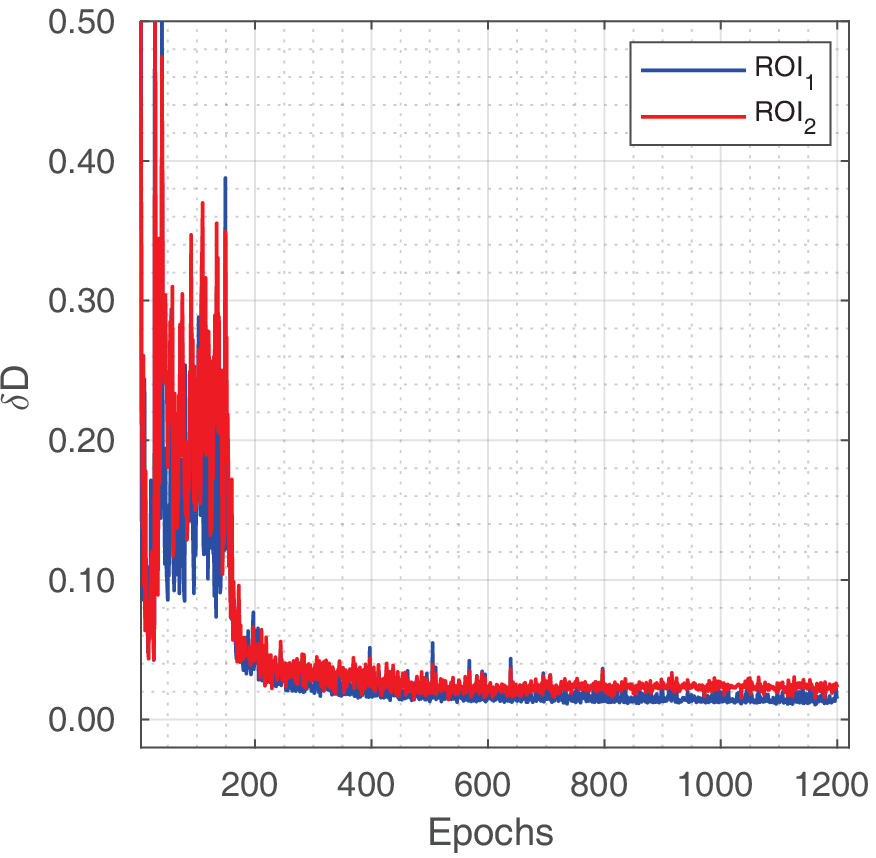}}
~\subfigure[]{\label{fig:BPShiftEpoch_c}\includegraphics[scale=.4]{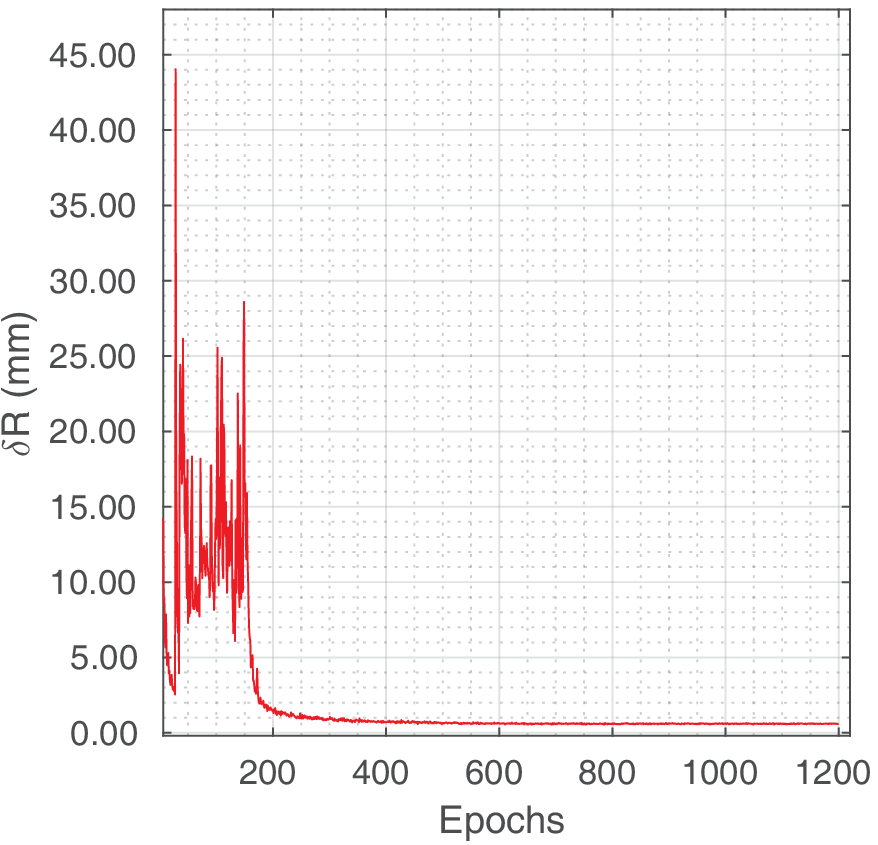}}
\end{center}
\caption{(a) MRE as an epoch function of ROI$_{1}$ and ROI$_{2}$ for the test dataset. (b) Absolute fractional dose difference as an epoch function of ROI$_{1}$ and ROI$_{2}$ for the test dataset. (c) BP shift as a function of epoch.}
\label{fig:DoseRangeEpochesab}
\end{figure}

During model training, we saved the trained model at each epoch to see how the model converged on the metric variables MRE, $\delta{D}$, and $\delta{R}$.   To obtain reasonably unbiased estimates, we examined the model performance on the test dataset after each epoch to estimate the stable phase of the model for up to 1200 epochs.
The stable values of the quality variables were derived from the 1D histograms for each epoch shown in Fig.~\ref{fig:DoseRangeEpochesab}.   The mean value of the histograms for each metric variable represents the model performance at that specific epoch.   Considerable convergence can be seen after 250 epochs for MRE, $\delta{D}$, and $\delta{R}$. The baseline prediction value of MRE is less than $2\%$, as shown in Fig.~\ref{fig:MRE_epoch_b}, and the error in $\delta{D}$ drops up to $3\%$ shown in Fig.~\ref{fig:FracDose_c} while the shift in BP is observed within $\pm 0.5\%$ as shown in Fig.~\ref{fig:BPShiftEpoch_c} after 250 epochs.   The uncertainty of the prediction in ROI$_{1}$ is smaller than ROI$_{2}$ at all epochs for the three metric variables. 
\begin{table}[ht!]
\caption{Summary of metrics for monoenergetic and SOBP   test datasets.} 
\centering % used for centering table
\begin{tabular}{c c c c} % centered columns (14 columns)
\hline\hline %inserts double horizontal lines
&&mono-energetic case& \\ [.5ex]
\hline
 & MRE & $\delta{D}$ & $\delta{R}$~(mm)\\ [1ex] % inserts table
%%heading
\hline % inserts single horizontal line
\hline
ROI$_1$ & 0.0162$\pm${0.0010}& 0.0218$\pm${0.0034}&  0.66$\pm${0.12}       \\\ 
ROI$_2$ & 0.0200$\pm${0.0013}&  0.0240$\pm${0.0039}&             \\[1ex]
\hline 
&&SOBP case& \\[.5ex]
\hline 
ROI$_1$ & 0.0187$\pm${0.0016}& 0.0226$\pm${0.0012}&  0.78$\pm${ 0.053}       \\\ 
ROI$_2$ & -0.0203$\pm${0.0017}&  0.0258$\pm${0.0014}&       \\[1ex]
\hline
\end{tabular}
\label{table:TestDataSummary} 
\end{table}

\subsection{Count-based analysis}
\label{subsection:Count_based_Analysis}
\begin{figure}[ht]
\begin{center}
\subfigure[]{\label{fig:MRECounts}\includegraphics[scale=.5]{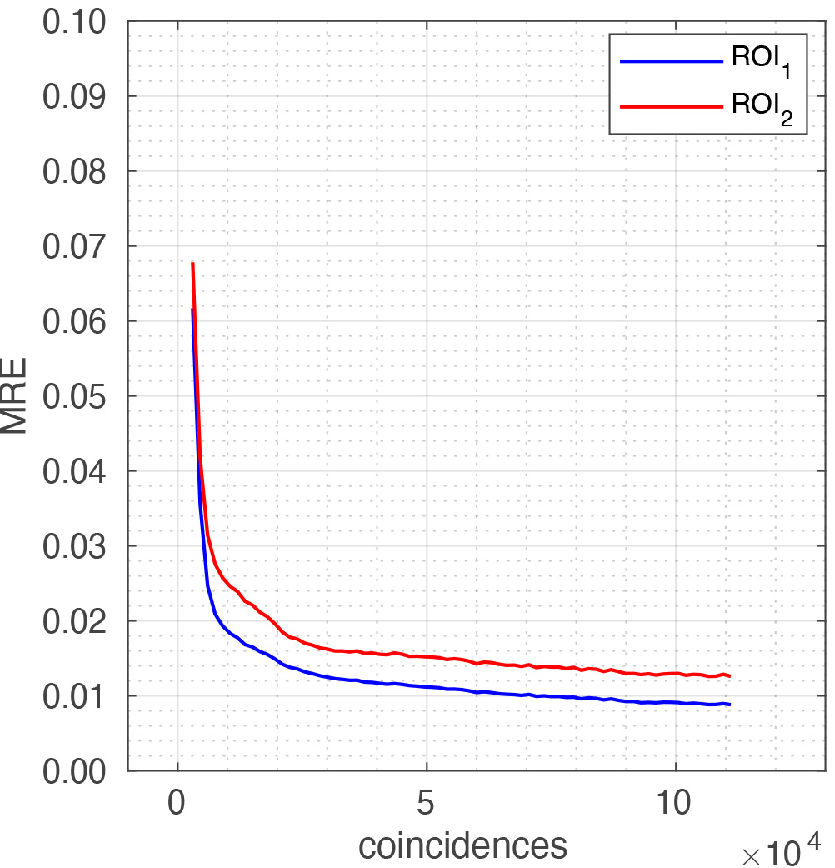}}
~\subfigure[]{\label{fig:abs_cou}\includegraphics[scale=.5]{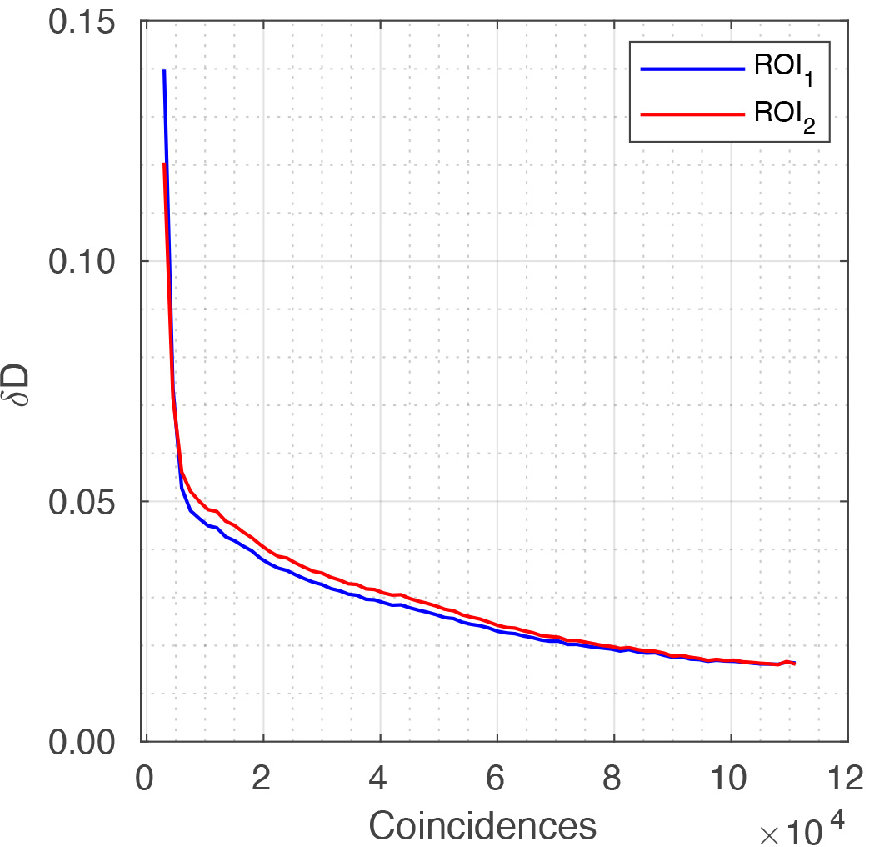}}
~\subfigure[]{\label{fig:ShiftInBPCOUNtBased}\includegraphics[scale=.5]{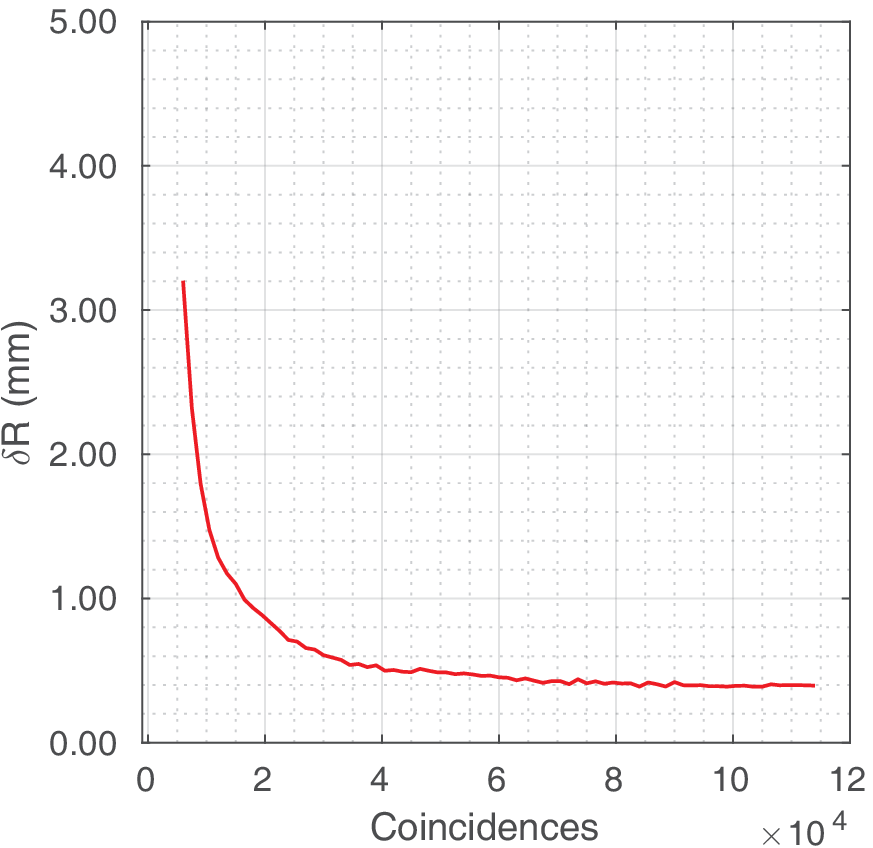}}
\subfigure[]{\label{fig:RMS_THRESHOLD_cnt}\includegraphics[scale=.45]{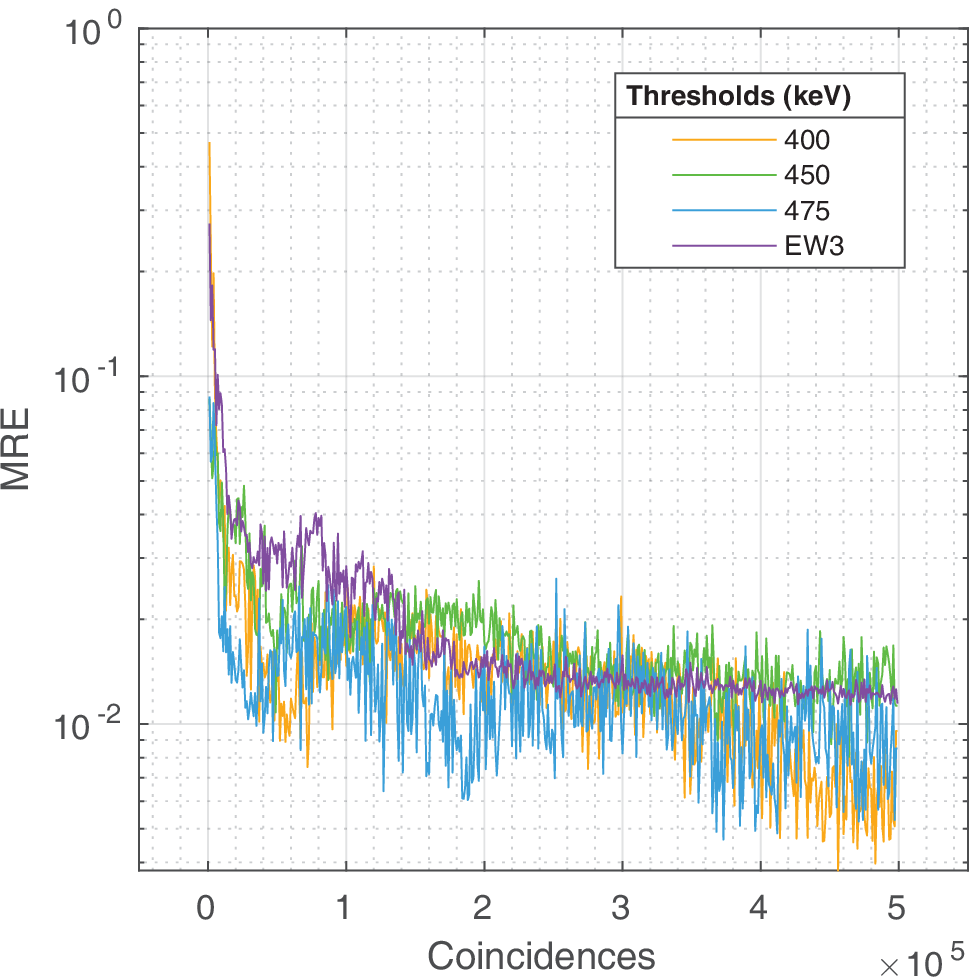}}
~\subfigure[]{\label{fig:ABS_THRESHOLD_cnt}\includegraphics[scale=.45]{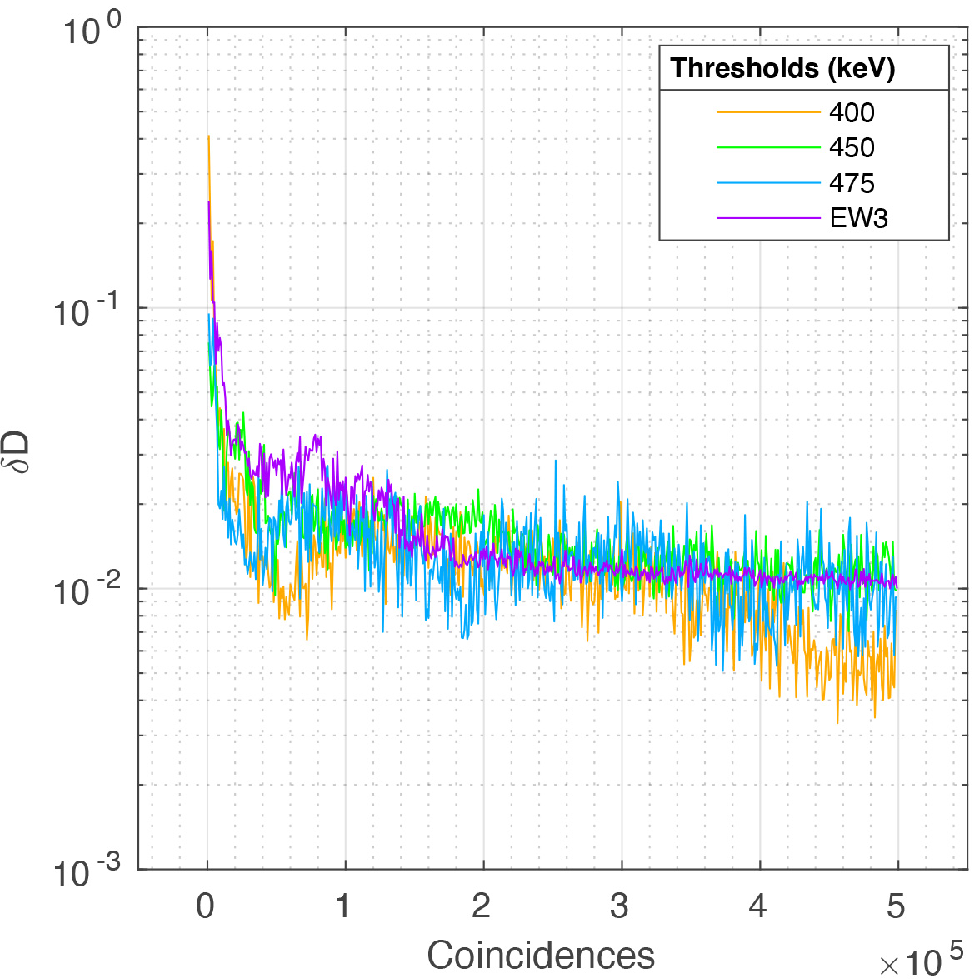}}
\end{center}
\caption{ (a)~MRE for ROI$_{1}$ and ROI$_{2}$ as a function of photopeak coincidences. (b) Absolute fractional dose difference for ROI$_{1}$ and ROI$_{2}$ as a function of photopeak coincidence data. (c) Shift in the BP as a function of photopeak coincidence data. (d) MRE as a function of coincidences for a 90 MeV proton beam on log$_{10}$ scale. (e) The fractional dose difference as a function of coincidences for a 90 MeV proton beam on log$_{10}$ scale.}
%\label{fig:DoseRangeCOUNTS}
\label{fig:countBasedStudy}
\end{figure}

The AI model was tested for a number of coincidence data to estimate the minimum number of coincidences on average to predict dose and BP shift. 
 The coincidence counts range from $7.3\times 10^3$ to $8\times 10^4$ for the 30 cm phantom case with the palm-sized detector described in Sec.~\ref{subsection:DetSimulation} with a scan time of 300 seconds shown in Fig.~\ref{fig:COIN_energy_a}.   To check the model performance with different numbers of coincidence events, we have performed training using approximately 10 times more coincidence data for all cases.
Fig.~\ref{fig:countBasedStudy} compares the accuracy of the model as a function of the coincidence data.   The number of coincidences detected in our palm-sized prototype detector can be improved by increasing the proton dose shown in Fig.~\ref{fig:ProtonFlux_Epectrum_a}.   In the count-based analysis, we calculated the average minimum number of coincidences for MRE, fractional dose, and shift in BP only using photopeak data with a threshold of 475 keV, as shown in Fig.~\ref{fig:countBasedStudy}(a)-(c).   It is worth noting that our photopeak map data range from $2\times10^{3}$ to $1.48\times10^{5}$  counts in a total of 74 bins with an increase of $2.0\times10^{3}$.   Table~\ref{table:SummaryCONTS2} summarizes the number of coincidences required for the prediction uncertainty of ROI$_{1}$ and ROI$_{2}$ to be less than 4\%, 3\%, 2\%, and 1\%.   Analysis for average coincidence was performed for only one threshold corresponding to the photopeak coincidences because preparing a complete test dataset of coincidence maps for all thresholds and energies requires extensively large disk space and analysis time.   But we have  shown the model performance for 90 MeV proton beam  as a function of the coincidences and important threshold conditions, as shown in Fig.~\ref{fig:countBasedStudy}(d)-(e). The model was trained  with $\times10^{6}$ photopeak coincidences for a particular case of 90 MeV proton beam case and the model requires $\sim10^{5}$ coincidences to achieve the uncertainty below 2\% in dose. 
\begin{table}[ht!]
\caption{A summary of the count-based study shows the number of photo-peak coincidences needed to achieve the level of uncertainty for $\delta{D}$, and $\delta{R}$.} 
\centering % used for centering table
\begin{tabular}{c 
c c c c } % centered columns (14 columns)
\hline\hline %inserts double horizontal lines
 Quantity  &  uncertainty  & \multicolumn{2}{c}{No. of photo peak coincidences}   &\\  
 \hline   
         &                  &    ROI$_1$  & ROI$_2$ \\
         
\hline 

%MRE & 4\% & 5660 & 6322  \\
%    & 3\% & 7019 & 8725 \\
%    & 2\% &  11143     &   25158    \\
$\delta{D}$ & 4\%& 23387& 23387\\
    & 3\% & 49098  & 57749  \\
    & 2\% &  98997     &  101770     \\
$\delta{R}$ &4\% &4749 & 4749\\
    & 3\% &  6436 &  6436 \\
    & 2\% &  11238     & 11238      \\
    & 1\% &  26911     & 26911      \\
\hline
\end{tabular}
\label{table:SummaryCONTS2} 
\end{table}
\subsection{Effect of detector threshold}
\begin{figure}[ht]
\centering     %%% not \center
\subfigure{\label{fig:MRE_Vth}\includegraphics[scale=.45]{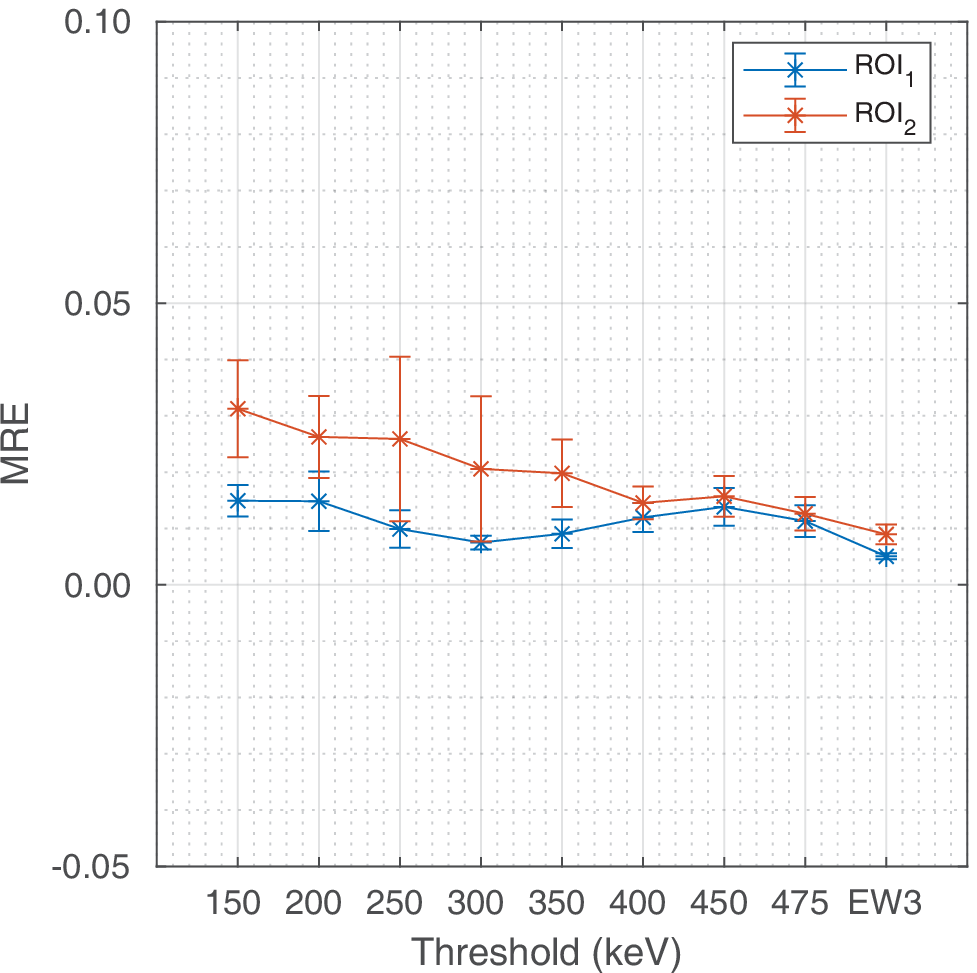}}
\subfigure{\label{fig:AbsDose_Vth}\includegraphics[scale=.45]{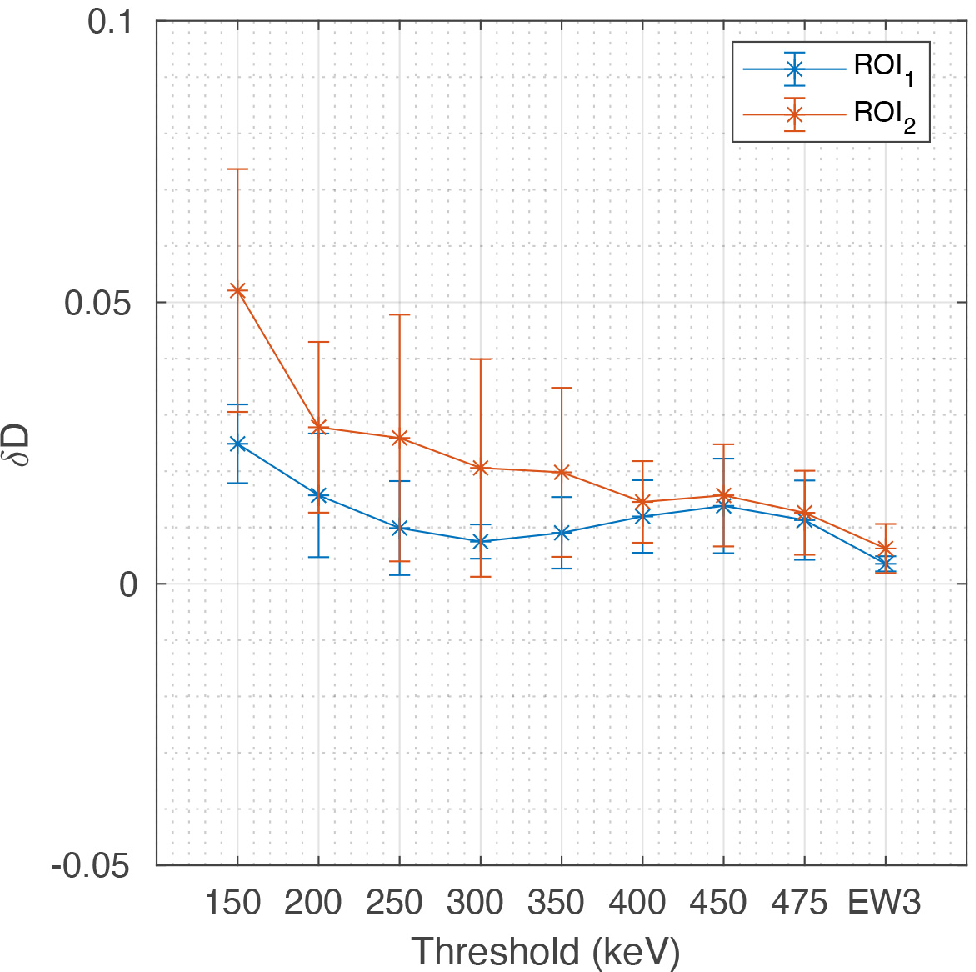}}
\subfigure{\label{fig:BPShift_Vth}\includegraphics[scale=.45]{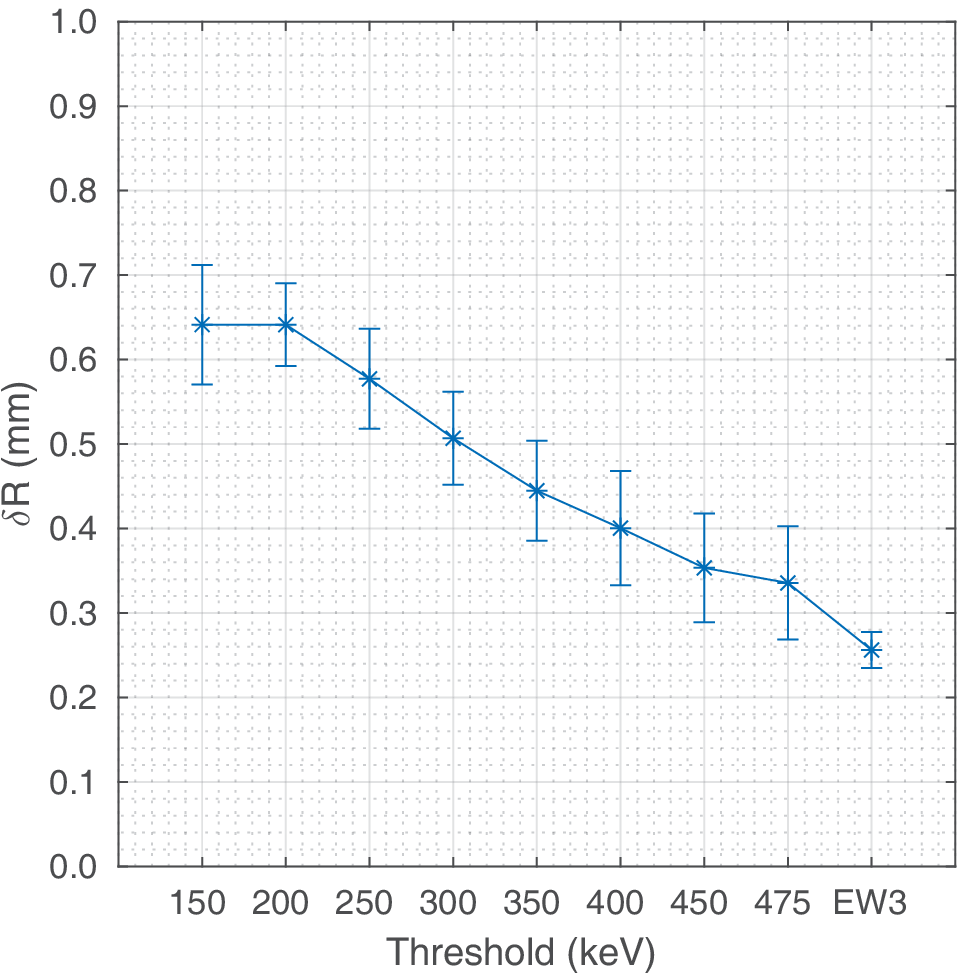}}
\caption{(a) MRE of ROI$_{1}$ and ROI$_{2}$ as a function of the detector threshold. (b) The absolute dose difference ($\delta{D}$) for ROI$_{1}$ and ROI$_{2}$ as a function of the detector threshold. (c) Shift in the BP as a function of the detector threshold.}
\label{fig:Metric_Vth}
\end{figure}

This section describes the performance of the model at different detector thresholds.   As shown in Fig.~\ref{fig:ProtonFlux_Epectrum_a}, the coincidence count rate is a function of the detector threshold.   The Compton peak ratio varies with the threshold, and the noise level can decrease significantly as the threshold increases.   We trained separately for each threshold and then measured the performance of the model using the test dataset reserved for each threshold.   The number of coincidences in the training datasets corresponding to each threshold is the same.   The number of photopeak coincidences increases with the threshold, but the total number of coincidences in each training dataset remains the same.   The prediction accuracy of the model increases as the threshold increases.   For the three-energy window (EW3) method, significantly low errors and stable values are observed, as shown in Fig.~\ref{fig:Metric_Vth}.   EW3 dataset includes three energy windows starting at a threshold of 150 KeV, as described in Sec.~\ref{subsection:DataPreparation}.

\subsection {Effect of cross-section  and material composition on model performance}
Our model is trained using coincidences data, and the effect of the cross-section can be deduced indirectly from the variation in the number of coincidences as the model is not directly sensitive to the cross-section changes but to coincidences. %A systematic uncertainty in cross-section will lead to a possible increase or decrease in the coincidence rate. 
A systematic uncertainty in cross-section may lead to a possible increase or decrease in the coincidence rate and variation in the local coincidence distributions. In our study, we are not taking  the local change in coincidence distribution into account but only the overall systematic increase or decrease in coincidences.
The behavior of coincidence change in Fig.~\ref{fig:ABS_THRESHOLD_cnt} gives a good idea that after training the model using 10$^6$ coincidences, the model is stable with good predictability up to 10 times smaller coincidences.%

\begin{figure}[ht]
\centering     %%% not \center
\subfigure[]{\label{fig:CT_variation_Dose}\includegraphics[scale=.45]{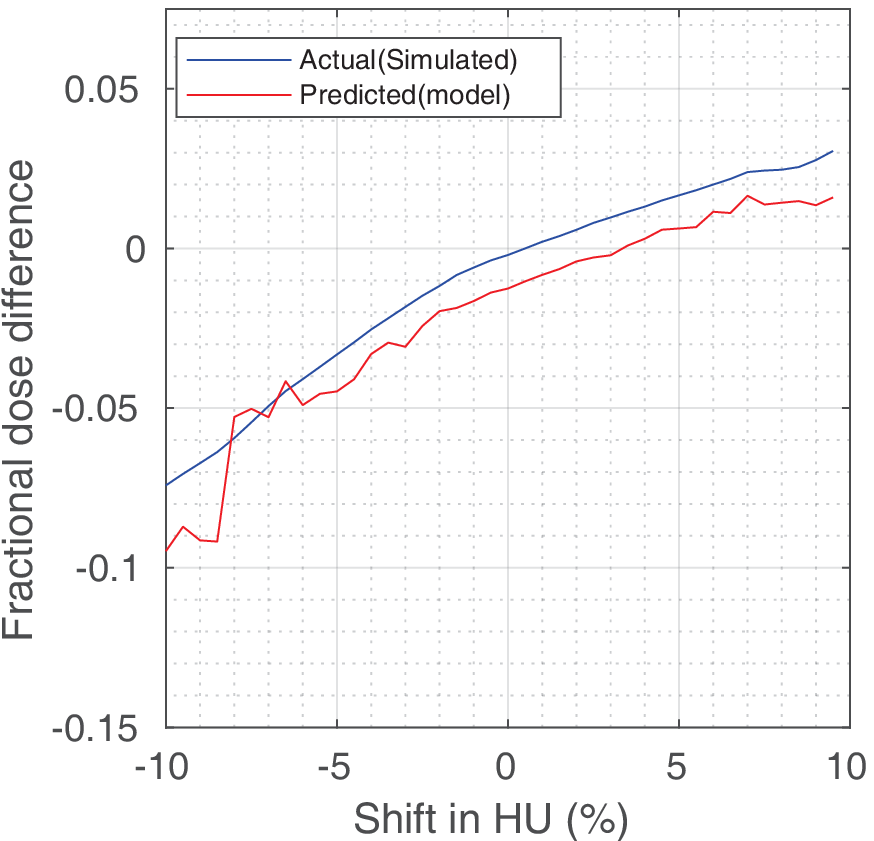}}
~\subfigure[]{\label{fig:CT_variation_BP}\includegraphics[scale=.45]{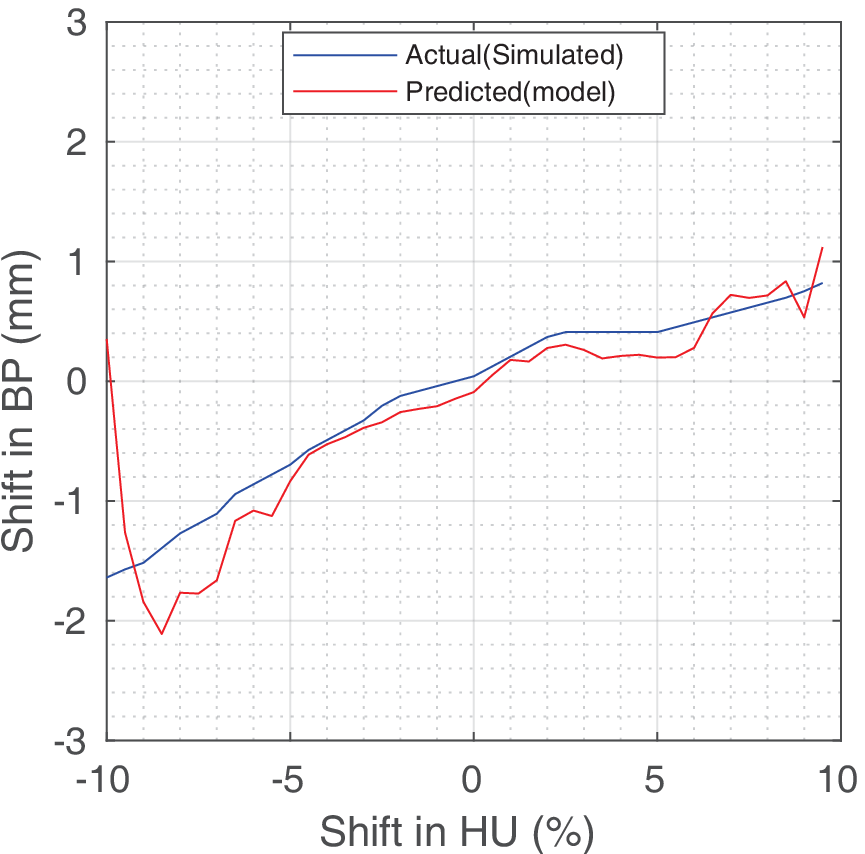}}
\caption{(a) The change in fractional dose with variation in HU values relative to dose of unmodified CT. (b) The change in BP positions with variation in HU values relative to the simulated BP position of the unmodified CT image. }
\label{fig:CT_Variations}
\end{figure}

The variation in the composition of the CT material was implemented by systematically changing the HU values of CT up to $\pm{10}$ from the original HU value of each voxel in the CT. The systematic change in CT material causes an overall change in the coincidence data but also introduces the changes in the local coincidence distribution. We calculate the change in dose and the change in BP for the systematically modified CT relative to the unmodified CT image. This change in dose and shift from BP is estimated for all varied CT images by using GATE simulations and is denoted by a blue curve. We also predict the change in dose and shift in BP relative to the simulated unmodified CT by using our model, which is denoted by the red curve in Fig~\ref{fig:CT_Variations}(a)-(b). The zero value on the $x$-axis corresponds to the results of the unmodified CT in Fig~\ref{fig:CT_Variations}. We found that the model predicts the 2\%  dose increase relative to the dose of unmodified CT for a 10\% systematic increase in HU values whereas for model predicts the 8.5\% decrease in dose for a 10\% systematic decrease in  CT values. Similarly, the model predicts that for $\pm{10}$\% change in CT values, the shift in BP varies from 1-2 mm change. The differences in the simulation and prediction from the model for dose and shift in BP are in agreement with the mean uncertainties of our model shown in Table~\ref{table:TestDataSummary}.

\subsection{Sensitivity consideration}
\begin{figure}[ht]
\centering     %%% not \center
\subfigure[]{\label{fig:sensitivity_DET}\includegraphics[scale=.4]{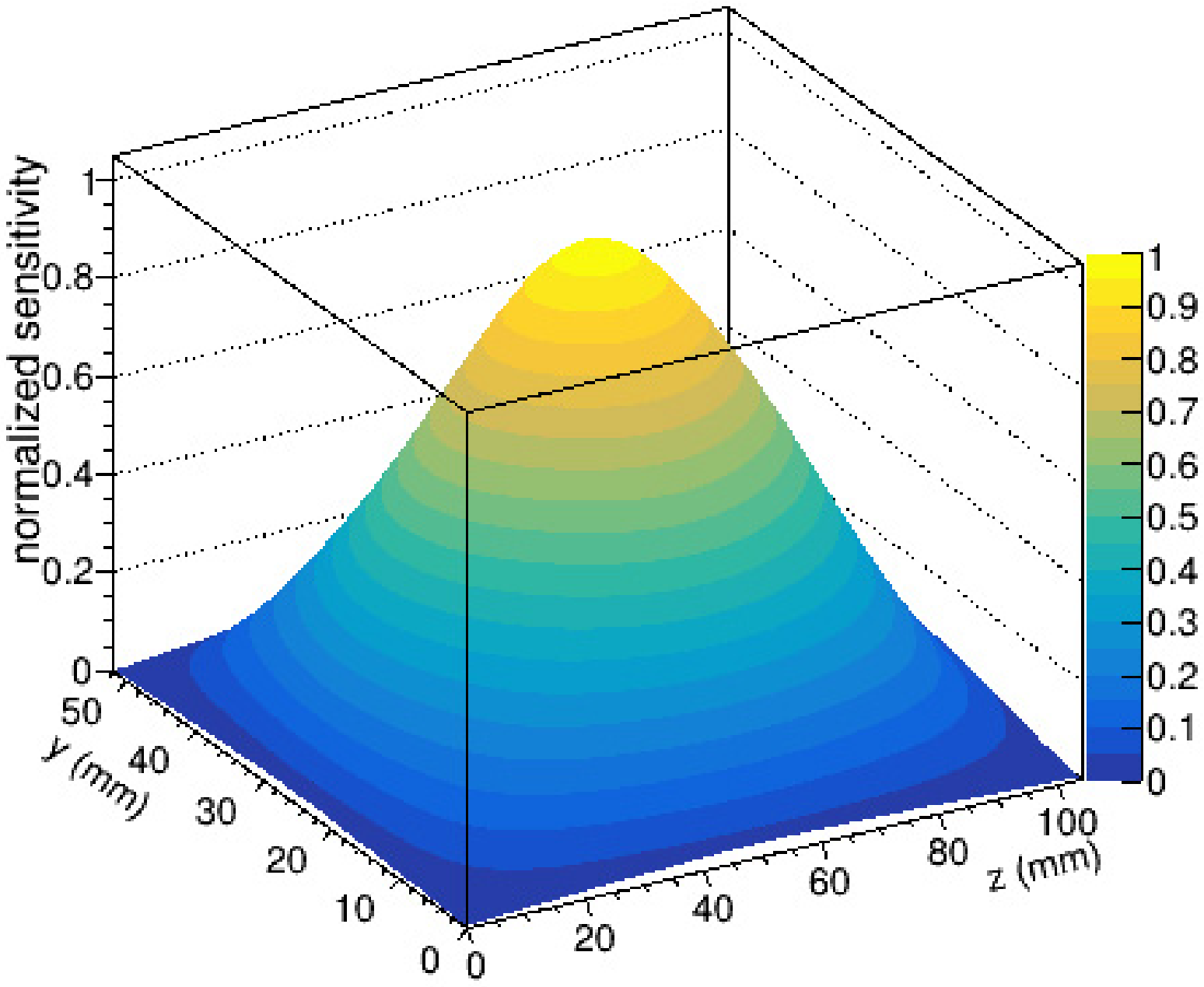}}
~~~~\subfigure[]{\label{fig:Coincidences_DET}\includegraphics[scale=.5]{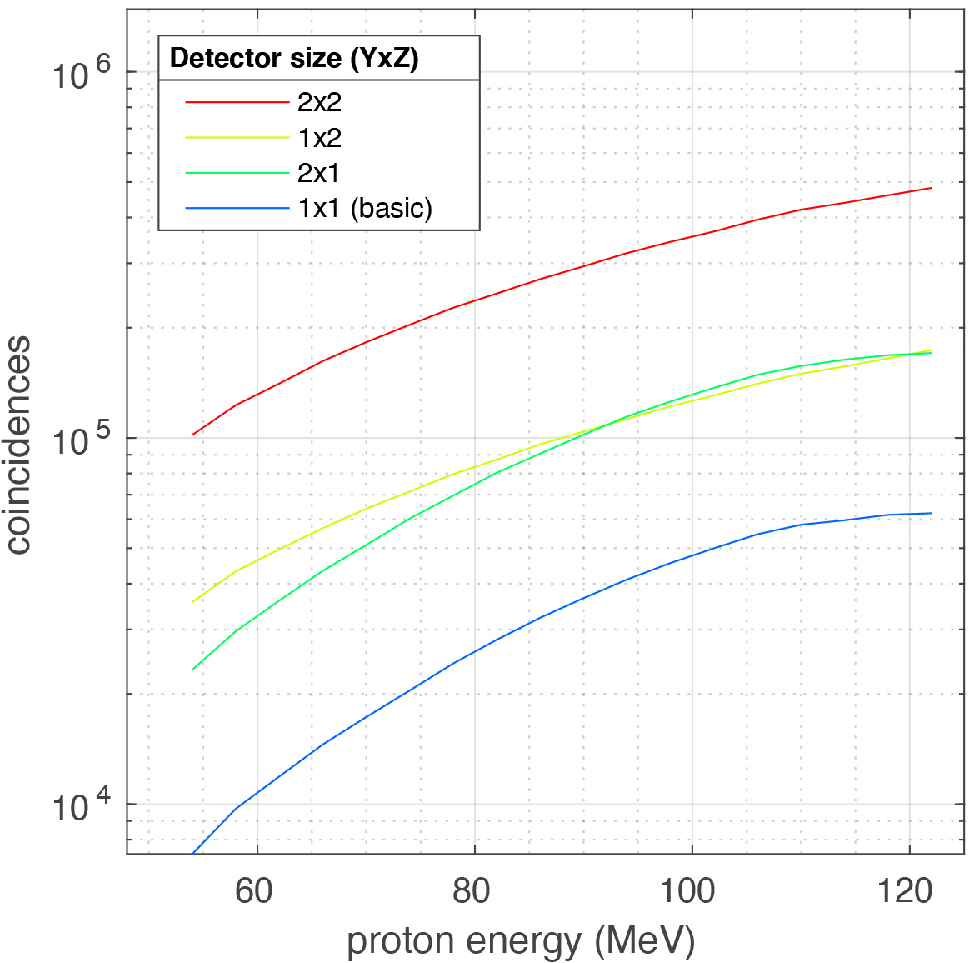}}
\caption{(a) The normalized sensitivity of the basic palm-sized PET module for 30~cm  phantom case, and (b) the effect on detected coincidences with increased detector sizes. }
\label{fig:COIN_Sensitivity}
\end{figure}

The count-based analysis in Sec.~\ref{subsection:Count_based_Analysis} shows that the performance of the model improves significantly with more coincidence data.   However, our detector prototype encountered low sensitivity, and due to the rectangular geometry of the detector, the absolute sensitivity in the $y$ and $z$ directions along the plane was not the same.   We calculated the absolute sensitivity by simulating a point source passing through the centerline in the $y$ and $z$ directions.   For our prototype, the absolute sensitivity at the mid-plane of the detector's field of view (FOV) is 0.91\%.   The off-center sensitivity is normalized to the maximum sensitivity position, which experiences a steep sensitivity loss, as shown in Fig.~\ref{fig:sensitivity_DET}. Increasing the area of the detection plan along the $y$ and $z$ directions can improve the sensitivity.   The increase in the detected coincidences for different proton energies and four detector sizes are shown in Fig.~\ref{fig:Coincidences_DET}.   By doubling the area of the base detector prototype, we can increase the detected coincidence by a factor of approximately 3.   By increasing the area by a factor of 4, the number of detected counts can be increased by a factor of 12.   To achieve the most stable prediction stage using the EW3 training method, about 10$^5$ coincidences are required, as shown in Fig.~\ref{fig:RMS_THRESHOLD_cnt}.   To achieve the 10$^5$ coincidence, a fourfold increase in detector geometry applies to the extreme case of 30~cm phantom size.   Another approach could be used to perform sensitivity correction or introduce some boosting algorithm to encrypt this information in the dataset passed to the deep learning model.

\section{Discussion}
%\begin{enumerate}
This work demonstrates the feasibility of directly mapping the proton dose from detected coincidences using raw data in less than one second (25~ms) using the trained model.  This method can bypass the conventional reconstruction chain and learn features directly from the projection data. 
We also demonstrate the feasibility of using our model to handle low count rates expected from compact detectors that are built to meet the tight mechanical requirements of proton therapy. Our dual-head setup has a geometrically active area 100 times smaller than the PET solution proposed by \citep{refe20}.

Generative adversarial networks face challenges in model convergence,  stability and mode collapse. Using diverse data for different energies, positions, and spot sizes helps to reduce the chance of model divergence and stability. Mode collapse is another severe problem for GANs models when the generator is not successful enough to produce sufficiently diverse data. Mode collapse is alleviated using cGANs where the model learns conditional probabilities using auxiliary information rather than random noise vectors from latent space.

In order to maximize the detected counts, it is recommended to center the detectors at the location of the Bragg peak in the depth dose distribution for a single energy and in the center of the SOBP for multiple energies. Larger detectors can provide more complete information on dose distribution.

To evaluate the performance of our method, we used more realistic and direct variables called mean relative error per pixel, fractional difference in absolute dose, and shift in BP. BP requires 5 to 9 times fewer coincidence data to converge faster than the data required for the dose to converge to the same level of accuracy. 
Using our detector prototype, the study proposes an estimated number of protons required for dose mapping.   According to Table~\ref{table:SummaryCONTS2}, approximately 23,000 coincidences are required to obtain absolute dose uncertainty below 4\% for ROI$_{1}$, which is obtained with  $2.2\times 10^{10}$ incident protons. This number is close to the single treatment fraction of $\sim$ 2Gy used to treat patients in multiple sessions.
To reach the 4\% prediction uncertainty in the BP position, the model needs only 47,000 photo-peak coincidences, which are equivalent to $3\times 10^{9}$ protons.   To approach an absolute dose difference of 2\%, approximately 5 times more photo-peak coincidences are required, which is equivalent to $1.5\times 10^{10}$ protons.   The lower the target uncertainty, the greater the number of coincidences. In the future, as the groups around the world are investigating hypo-fractionated therapy and flash irradiation ($>30$ Gys$^{-1}$ \citep{verhaegen2021considerations} higher proton fluxes and therefore, higher coincidences can be expected, which will lead to lower uncertainties on the presented model output. The usage of PET detectors becomes especially relevant in comparison to the prompt gamma detectors, where it is challenging to acquire a large amount of data in a sub-second time frame.

The uncertainties on the CT values can impact the stopping power and hence the dose distribution, it can also lead to a mismatch in the material assignment and hence to a different distribution of the positron emitters. 
\citep{EspagnaPaganetti2010} show that the uncertainties in the proton range are below 0.5 mm ($\sim$ 0.2\%) when the CT conversion used by Monte Carlo is based on stoichiometric techniques such as prescribed by \citep{refe31}. Materials such as lungs can have large inhomogeneity and are more challenging to achieve low uncertainties. 
In our conversion scheme, the variation of the dominant positron emitters for materials for increasing values of HU in our conversion table is smooth and remains under 10\% (except lung). The actual variation in the coincidences for a $\pm{10}$ HU shift can be expected to be under 2\% where the model remains stable.
It is recommended that our model be applied to homogeneous materials to achieve better reliability. The model predictions can deviate from the training data significantly for heterogeneous materials such as the lungs due to the possibility of averaging the HU of tissues with a large difference in density values.

 The choice of centering the phantom in the detector FOV is based on maximizing the counts and is not an absolute requirement. For cases where off-center positioning is required, the model will have to be trained on that geometry, enabling it to learn the features of off-center-induced asymmetries in the coincidence maps. This is not a limitation of the current model. Furthermore, the model is able to remain stable for small deviations in the positioning. For small offsets less than 10 mm in the irradiation positions, our current model predicts the dose with uncertainty less than 2.5\% and a shift in BP below 1 mm.\\
One potential limitation of the present model is the requirement of time and resources for training purposes. In this work, we have trained the model on various phantom positions and beam energies to demonstrate the robustness of the model. However, for practical application of this model, we recommend training the model in the exact case of the planned energies of treatment at the planned tumor locations. This makes training significantly faster and more reliable.
We demonstrate that the provision of additional information on photon energy improves the model accuracy. Using multiple channels in the input condition is a superior approach because it increases the data conditions. Hence, more kernel channels are available for convolution, summed up as a single cross-correlation map. Although photo-peak data are higher in quality, the required counts are 8 times higher to compete with the same level of accuracy, which is achieved using a multi-channel approach. In the future other information pertaining to the beam delivery, the patient DICOM, the photon energy, time-of-flight information measured, etc should be incorporated as prior information to increase the scope of the model.
  
\section {Conclusion}
 This work demonstrates the potential for 2D dose validation and direct reconstruction using compact palm-sized PET modules using the potential of the cGAN-based deep learning framework.   In our study, mean relative error, absolute dose difference, and shift in BP were used as figures of merit.   Our method can obtain Bragg peak position results with less than 1\% and absolute dose with less than 2\% uncertainty achieved with $\sim 10^5$ coincidences when irradiating with mono-energetic proton beams between 50 MeV and 120 MeV for less than five minutes of acquisition.   Our study opens a window for exploring the application of deep learning models in dose mapping using small amounts of coincidence data. 
 Including auxiliary information corresponding to the detected photon energy by feeding different conditions related to different set thresholds leads to improved stability in the predicted output.   The fractional dose difference between the predicted dose and the true dose distribution improves from 2.6\% to 1.6\% by providing the additional condition corresponding to the photon energy. It is presently recommended that the model be trained for each patient specifically geared to the planned treatment to achieve reliable training in a short time. Including additional information on the material and the beam delivery conditions, and the time and energy information of the individual signals as prior information to the neural network model can further enhance the scope of this model. Our model can be a valuable tool for treatment planning and quality assurance in particle therapy.
\section*{Acknowledgement}
The research and development were funded by the Academia Sinica thematic research program with the project number AS-TP-108-ML0.   This work also used ASGC (Academia Sinica Grid-Computing Center) Distributed Cloud resources, which is supported by Academia Sinica, Taiwan.
\newpage
\newcommand{\newblock}{}
\bibliographystyle{agsm} 
%\bibliography{Biblo}

@article{refe1,
title = {Range uncertainties in proton therapy and the role of Monte Carlo simulations},
doi = {10.1088/0031-9155/57/11/r99},
url = {https://doi.org/10.1088/0031-9155/57/11/r99},
year = 2012,
month = {may},
publisher = {{IOP} Publishing},
volume = {57},
number = {11},
pages = {R99--R117},
author = {Harald Paganetti}
}

@article{refe2,
title = {Potential application of {PET} in quality assurance of proton therapy},
doi = {10.1088/0031-9155/45/11/403},
url = {https://doi.org/10.1088/0031-9155/45/11/403},
year = 2000,
month = {oct},
publisher = {{IOP} Publishing},
volume = {45},
number = {11},
pages = {N151--N156},
author = {K Parodi}
}

@article{refe3,
title = {Patient Study of In Vivo Verification of Beam Delivery and Range, Using Positron Emission Tomography and Computed Tomography Imaging After Proton Therapy},
journal = {International Journal of Radiation Oncology Biology Physics},
volume = {68},
number = {3},
pages = {920-934},
year = {2007},
issn = {0360-3016},
doi = {https://doi.org/10.1016/j.ijrobp.2007.01.063},
url = {https://www.sciencedirect.com/science/article/pii/S036030160700377X},
author = {Katia Parodi and Harald Paganetti and Helen A. Shih and Susan Michaud and Jay S. Loeffler and Thomas F. DeLaney and Norbert J. Liebsch and John E. Munzenrider and Alan J. Fischman and Antje Knopf and Thomas Bortfeld}
}

@article{min2006prompt,
  title={Prompt gamma measurements for locating the dose falloff region in the proton therapy},
  author={Min, Chul-Hee and Kim, Chan Hyeong and Youn, Min-Young and Kim, Jong-Won},
  journal={Applied physics letters},
  volume={89},
  number={18},
  pages={183517},
  year={2006},
  publisher={American Institute of Physics}
}

@ARTICLE{refe4,
  author={Paans, A.M.J. and Schippers, J.M.},
  journal={IEEE Transactions on Nuclear Science}, 
  title={Proton therapy in combination with PET as monitor: a feasibility study}, 
  year={1993},
  volume={40},
  number={4},
  pages={1041-1044},
  doi={10.1109/23.256709}
  }

@ARTICLE{refe5,
  author={Nemallapudi, Mythra Varun and Rahman, Atiq and Chen, Augustine Ei-Fong and Lee, Shih-Chang and Lin, Chih-Hsun and Chu, Ming-Lee and Chou, Chen-Ying},
  journal={IEEE Transactions on Radiation and Plasma Medical Sciences}, 
  title={Positron emitter depth distribution in PMMA irradiated with 130 MeV protons measured using TOF-PET detectors.}, 
  year={2021},
  volume={},
  number={},
  pages={1-1},
  doi={10.1109/TRPMS.2021.3084953}
  }

@article{refe5_2,
  title={Feasibility of quasi-prompt PET-based range verification in proton therapy},
  author={Ozoemelam, Ikechi and Van der Graaf, Emiel and Van Goethem, Marc-Jan and Kapusta, Maciej and Zhang, Nan and Brandenburg, Sytze and Dendooven, Peter},
  journal={Physics in Medicine \& Biology},
  volume={65},
  number={24},
  pages={245013},
  year={2020},
  publisher={IOP Publishing}
}

@article{refe7,
author = {Vaquero, Juan José and Kinahan, Paul},
title = {Positron Emission Tomography: Current Challenges and Opportunities for Technological Advances in Clinical and Preclinical Imaging Systems},
journal = {Annual Review of Biomedical Engineering},
volume = {17},
number = {1},
pages = {385-414},
year = {2015},
doi = {10.1146/annurev-bioeng-071114-040723},
    note ={PMID: 26643024},
URL = {     https://doi.org/10.1146/annurev-bioeng-071114-040723},
eprint = { https://doi.org/10.1146/annurev-bioeng-071114-040723 
}}

@article{alessio2005analytical,
  title={Analytical reconstruction of deconvolved Fourier rebinned PET sinograms},
  author={Alessio, Adam and Sauer, Ken and Kinahan, Paul},
  journal={Physics in medicine \& biology},
  volume={51},
  number={1},
  pages={77},
  year={2005},
  publisher={IOP Publishing}
}

@article{lu2008,
  title={A potential method for in vivo range verification in proton therapy treatment},
  author={Lu, Hsiao-Ming},
  journal={Physics in Medicine \& Biology},
  volume={53},
  number={5},
  pages={1413},
  year={2008},
  publisher={IOP Publishing}
}

@inproceedings{watts2009proton,
  title={A proton range telescope for quality assurance in hadrontherapy},
  author={Watts, David A and Amaldi, Ugo and Go, Apollo and Hajdas, Wojtek and Iliescu, Sarolta and Malakhov, Nail and Samarati, Jerome and Sauli, Fabio and others},
  booktitle={2009 IEEE Nuclear Science Symposium Conference Record (NSS/MIC)},
  pages={4163--4166},
  year={2009},
  organization={IEEE}
}

@article {refe8,
	Title = {Image reconstruction for PET/CT scanners: past achievements and future challenges},
	Author = {Tong, Shan and Alessio, Adam M and Kinahan, Paul E},
	DOI = {10.2217/iim.10.49},
	Number = {5},
	Volume = {2},
	Month = {October},
	Year = {2010},
	Journal = {Imaging in medicine},
	ISSN = {1755-5191},
	Pages = {529—545}
	}

@article{refe9,
title = {System models for PET statistical iterative reconstruction: A review},
journal = {Computerized Medical Imaging and Graphics},
volume = {48},
pages = {30-48},
year = {2016},
issn = {0895-6111},
doi = {https://doi.org/10.1016/j.compmedimag.2015.12.003},
url = {https://www.sciencedirect.com/science/article/pii/S0895611115001901},
author = {A. Iriarte and R. Marabini and S. Matej and C.O.S. Sorzano and R.M. Lewitt}
}

@article{refe10,
title = {Dose quantification from in-beam positron emission tomography},
journal = {Radiotherapy and Oncology},
volume = {73},
pages = {S96-S98},
year = {2004},
note = {Carbon-Ion Theraphy},
issn = {0167-8140},
doi = {https://doi.org/10.1016/S0167-8140(04)80024-0},
url = {https://www.sciencedirect.com/science/article/pii/S0167814004800240},
author = {W. Enghardt and K. Parodi and P. Crespo and F. Fiedler and J. Pawelke and F. Pönisch}
}

@article{refe10_1,
  title={Monitoring proton therapy with PET},
  author={Paganetti, H and El Fakhri, G},
  journal={The British journal of radiology},
  volume={88},
  number={1051},
  pages={20150173},
  year={2015},
  publisher={The British Institute of Radiology.}
}

@article{refe11,
	doi = {10.1088/0031-9155/54/11/n02},
	url = {https://doi.org/10.1088/0031-9155/54/11/n02},
	year = 2009,
	month = {may},
	publisher = {{IOP} Publishing},
	volume = {54},
	number = {11},
	pages = {N217--N228},
	author = {E Fourkal and J Fan and I Veltchev},
	title = {Absolute dose reconstruction in proton therapy using {PET} imaging modality: feasibility study}
	}

@article{refe12,
  title={On the effectiveness of ion range determination from in-beam PET data},
  author={Fiedler, Fine and Shakirin, Georgy and Skowron, Judith and Braess, Henning and Crespo, Paulo and Kunath, Daniela and Pawelke, J{\"o}rg and P{\"o}nisch, Falk and Enghardt, Wolfgang},
  journal={Physics in Medicine \& Biology},
  volume={55},
  number={7},
  pages={1989},
  year={2010},
  publisher={IOP Publishing}
}

@article{refe14,
	doi = {10.1088/0031-9155/51/8/003},
	url = {https://doi.org/10.1088/0031-9155/51/8/003},
	year = 2006,
	month = {mar},
	publisher = {{IOP} Publishing},
	volume = {51},
	number = {8},
	pages = {1991--2009},
	author = {Katia Parodi and Thomas Bortfeld},
	title = {A filtering approach based on Gaussian{\textendash}powerlaw convolutions for local {PET} verification of proton radiotherapy}
	}

@article{refe_m,
  title={Data acquisition in PET imaging},
  author={Fahey, Frederic H},
  journal={Journal of nuclear medicine technology},
  volume={30},
  number={2},
  pages={39--49},
  year={2002},
  publisher={Soc Nuclear Med}
}

@article{refe15,
	doi = {10.1088/0031-9155/56/16/001},
	url = {https://doi.org/10.1088/0031-9155/56/16/001},
	year = 2011,
	month = {jul},
	publisher = {{IOP} Publishing},
	volume = {56},
	number = {16},
	pages = {5079--5098},
	author = {F Attanasi and A Knopf and K Parodi and H Paganetti and T Bortfeld and V Rosso and A Del Guerra},
	title = {Extension and validation of an analytical model for in-vivo PET verification of proton therapy{\textemdash}a phantom and clinical study}
	}

@article{refe16,
	doi = {10.1088/0031-9155/56/23/017},
	url = {https://doi.org/10.1088/0031-9155/56/23/017},
	year = 2011,
	month = {nov},
	publisher = {{IOP} Publishing},
	volume = {56},
	number = {23},
	pages = {7601--7619},
	author = {Steffen Remmele and Jürgen Hesser and Harald Paganetti and Thomas Bortfeld},
	title = {A deconvolution approach for {PET}-based dose reconstruction in proton radiotherapy}
	}

@article{refe18,
	doi = {10.1088/1361-6560/ab3276},
	url = {https://doi.org/10.1088/1361-6560/ab3276},
	year = 2019,
	month = {sep},
	publisher = {{IOP} Publishing},
	volume = {64},
	number = {17},
	pages = {175011},
	author = {Takamitsu Masuda and Teiji Nishio and Jun Kataoka and Makoto Arimoto and Akira Sano and Kumiko Karasawa},
	title = {{ML}-{EM} algorithm for dose estimation using {PET} in proton therapy}
	}

@article{refe19,
	doi = {10.1088/1361-6560/ab3564},
	url = {https://doi.org/10.1088/1361-6560/ab3564},
	year = 2019,
	month = {sep},
	publisher = {{IOP} Publishing},
	volume = {64},
	number = {17},
	pages = {175009},
	author = {Chuang Liu and Zhongxing Li and Wenbin Hu and Lei Xing and Hao Peng},
	title = {Range and dose verification in proton therapy using proton-induced positron emitters and recurrent neural networks ({RNNs})}
	}

@article{refe20,
	doi = {10.1088/1361-6560/ab9707},
	url = {https://doi.org/10.1088/1361-6560/ab9707},
	year = 2020,
	month = {sep},
	publisher = {{IOP} Publishing},
	volume = {65},
	number = {18},
	pages = {185003},
	author = {Zongsheng Hu and Guangyao Li and Xiaoke Zhang and Kuangkuang Ye and Jiade Lu and Hao Peng},
	title = {A machine learning framework with anatomical prior for online dose verification using positron emitters and {PET} in proton therapy}
	}

@article{refe20_1,
  title={Dose calculation in proton therapy using a discovery cross-domain generative adversarial network (DiscoGAN)},
  author={Zhang, Xiaoke and Hu, Zongsheng and Zhang, Guoliang and Zhuang, Yongdong and Wang, Yuenan and Peng, Hao},
  journal={Medical Physics},
  volume={48},
  number={5},
  pages={2646--2660},
  year={2021},
  publisher={Wiley Online Library}
}

@InProceedings{refe22,
author={Ronneberger, Olaf and Fischer, Philippand Brox, Thomas},
editor={Navab, Nassir
and Hornegger, Joachim
and Wells, William M.
and Frangi, Alejandro F.},
title={U-Net: Convolutional Networks for Biomedical Image Segmentation},
booktitle={Medical Image Computing and Computer-Assisted Intervention -- MICCAI 2015},
year={2015},
publisher={Springer International Publishing},
address="Cham",
pages="234--241",
isbn="978-3-319-24574-4"
}

@article{refe23,
  title={Image-to-Image Translation with Conditional Adversarial Networks},
  author={Phillip Isola and Jun-Yan Zhu and Tinghui Zhou and Alexei A. Efros},
  journal={2017 IEEE Conference on Computer Vision and Pattern Recognition (CVPR)},
  year={2017},
  pages={5967-5976}
}

@inproceedings{oussidi2018deep,
  title={Deep generative models: Survey},
  author={Oussidi, Achraf and Elhassouny, Azeddine},
  booktitle={2018 International Conference on Intelligent Systems and Computer Vision (ISCV)},
  pages={1--8},
  year={2018},
  organization={IEEE}
}

@article{refe24,
title = {DeepPET: A deep encoder–decoder network for directly solving the PET image reconstruction inverse problem},
journal = {Medical Image Analysis},
volume = {54},
pages = {253-262},
year = {2019},
issn = {1361-8415},
doi = {https://doi.org/10.1016/j.media.2019.03.013},
url = {https://www.sciencedirect.com/science/article/pii/S1361841518305838},
author = {Ida Häggström and C. Ross Schmidtlein and Gabriele Campanella and Thomas J. Fuchs}
}

@article{refe26,
	doi = {10.1088/0031-9155/56/4/001},
	url = {https://doi.org/10.1088/0031-9155/56/4/001},
	year = 2011,
	month = {jan},
	publisher = {{IOP} Publishing},
	volume = {56},
	number = {4},
	pages = {881--901},
	author = {S Jan and D Benoit and E Becheva and T Carlier and F Cassol and P Descourt and T Frisson and L Grevillot and L Guigues and L Maigne and C Morel and Y Perrot and N Rehfeld and D Sarrut and D R Schaart and S Stute and U Pietrzyk and D Visvikis and N Zahra and I Buvat},
	title = {{GATE} V6: a major enhancement of the {GATE} simulation platform enabling modelling of {CT} and radiotherapy}
	}

@article{refe27,
	doi = {10.1088/0031-9155/56/16/008},
	url = {https://doi.org/10.1088/0031-9155/56/16/008},
	year = 2011,
	month = {jul},
	publisher = {{IOP} Publishing},
	volume = {56},
	number = {16},
	pages = {5203--5219},
	author = {L Grevillot and D Bertrand and F Dessy and N Freud and D Sarrut},
	title = {A Monte Carlo pencil beam scanning model for proton treatment plan simulation using {GATE}/{GEANT}4}
	}

@article{MIRZA,
  title={Conditional generative adversarial nets},
  author={Mirza, Mehdi and Osindero, Simon},
  journal={arXiv preprint arXiv:1411.1784},
  year={2014}
}

@article{refe28,
  title={Evaluation of GATE-RTion (GATE/Geant4) Monte Carlo simulation settings for proton pencil beam scanning quality assurance},
  author={Winterhalter, Carla and Taylor, Michael and Boersma, David and Elia, Alessio and Guatelli, Susanna and Mackay, Ranald and Kirkby, Karen and Maigne, Lydia and Ivanchenko, Vladimir and Resch, Andreas F and others},
  journal={Medical Physics},
  volume={47},
  number={11},
  pages={5817--5828},
  year={2020},
  publisher={Wiley Online Library}
  }

@article{refe31,
	doi = {10.1088/0031-9155/45/2/314},
	url = {https://doi.org/10.1088/0031-9155/45/2/314},
	year = 2000,
	month = {jan},
	publisher = {{IOP} Publishing},
	volume = {45},
	number = {2},
	pages = {459--478},
	author = {Wilfried Schneider and Thomas Bortfeld and Wolfgang Schlegel},
	title = {Correlation between {CT} numbers and tissue parameters needed for Monte Carlo simulations of clinical dose distributions}
}

@article{verhaegen2021considerations,
  title={Considerations for shoot-through FLASH proton therapy},
  author={Verhaegen, Frank and Wanders, Roel-Germ and Wolfs, Cecile and Eekers, Dani{\"e}lle},
  journal={Physics in Medicine \& Biology},
  volume={66},
  number={6},
  pages={06NT01},
  year={2021},
  publisher={IOP Publishing}
}

@article{jette2011creating,
  title={Creating a spread-out Bragg peak in proton beams},
  author={Jette, David and Chen, Weimin},
  journal={Physics in Medicine \& Biology},
  volume={56},
  number={11},
  pages={N131},
  year={2011},
  publisher={IOP Publishing}
}

@article{bortfeld1996analytical,
  title={An analytical approximation of depth-dose distributions for therapeutic proton beams},
  author={Bortfeld, Thomas and Schlegel, Wolfgang},
  journal={Physics in Medicine \& Biology},
  volume={41},
  number={8},
  pages={1331},
  year={1996},
  publisher={IOP Publishing}
}

@incollection{Asako,
  title={Deep learning for multimodal data fusion},
  author={Kanezaki, Asako and Kuga, Ryohei and Sugano, Yusuke and Matsushita, Yasuyuki},
  booktitle={Multimodal Scene Understanding},
  pages={20--22},
  year={2019},
  publisher={Elsevier}
}

@inproceedings{l1-lossfunction,
 author = {Mukherjee, Subhadip and Carioni, Marcello and \"{O}ktem, Ozan and Sch\"{o}nlieb, Carola-Bibiane},
 booktitle = {Advances in Neural Information Processing Systems},
 editor = {M. Ranzato and A. Beygelzimer and Y. Dauphin and P.S. Liang and J. Wortman Vaughan},
 pages = {21413--21425},
 publisher = {Curran Associates, Inc.},
 title = {End-to-end reconstruction meets data-driven regularization for inverse problems},
 url = {https://proceedings.neurips.cc/paper/2021/file/b2df0a0d4116c55f81fd5aa1ef876510-Paper.pdf},
 volume = {34},
 year = {2021}
}

@article{EspagnaPaganetti2010, 
title = {The impact of uncertainties in the CT conversion algorithm when predicting proton beam ranges in patients from dose and PET-activity distributions},
doi = {10.1088/0031-9155/55/24/011},
year = 2010,
journal={Physics in Medicine \& Biology},
volume = {55},
number = {},
pages = {7557-7571},
author={{Espagna S. and Paganetti H.}}
}

@InProceedings{Lin_2018_CVPR,
author = {Lin, Jianxin and Xia, Yingce and Qin, Tao and Chen, Zhibo and Liu, Tie-Yan},
title = {Conditional Image-to-Image Translation},
booktitle = {Proceedings of the IEEE Conference on Computer Vision and Pattern Recognition (CVPR)},
month = {June},
year = {2018}
}

\end{document}